\documentclass[12pt]{iopart}
\usepackage[legalpaper,margin=1in]{geometry}
\usepackage[english]{babel}
\usepackage[colorlinks,allcolors=blue]{hyperref}
\usepackage[linesnumbered]{algorithm2e}
\usepackage{amsmath}
\usepackage{amsfonts}
\usepackage{amssymb}
\usepackage{amsthm}
\usepackage{enumitem}
\usepackage{mathtools}
\usepackage{multirow}
\usepackage{subcaption}
\usepackage{xcolor}

\usepackage{natbib}
\theoremstyle{definition}
\newtheorem{definition}{Definition}
\newtheorem{assump}{Assumption}
\theoremstyle{plain}
\newtheorem{theorem}{Theorem}
\newtheorem{rem}{Remark}

\DeclareMathOperator*{\argmin}{arg\,min}
\newcommand{\R}{\mathbb{R}}
\newcommand{\bR}{\mathcal{R}}
\newcommand{\Y}{\mathcal{Y}}
\newcommand{\F}{\mathcal{F}}
\newcommand{\M}{\mathcal{M}}
\renewcommand{\P}{\mathbb{P}}
\newcommand{\Q}{\mathbb{Q}}
\newcommand{\E}{\mathbb{E}}

\newcommand{\y}{\boldsymbol{y}}
\newcommand{\ynew}{y^{new}}
\newcommand{\hrho}{\hat{\rho}}
\newcommand{\loo}[1]{#1_{(-i)}}
\newcommand{\kl}{D_{KL}}

\RestyleAlgo{ruled}
\SetKwComment{Comment}{/* }{ */}

\begin{document}
\title{
    Generalized Bayes Approach to Inverse Problems with Model Misspecification
}
\author{
    Youngsoo Baek$^1$,
    Wilkins Aquino$^2$, 
    Sayan Mukherjee$^{1~3~4~5}$
}
\address{$^1$Department of Statistical Science, Duke University, NC}
\address{Department of Mechanical Engineering and Materials Science, Duke University, NC}
\address{$^3$ Department of Mathematics, Computer Science, Biostatistics \& Bioinformatics, Duke University, NC}
\address{Center for Scalable Data Analytics and Artificial Intelligence, Universit\"at Leipzig}
\address{Max Planck Institute for Mathematics in the Sciences, Leipzig}

\begin{abstract}
We propose a general framework for obtaining probabilistic solutions to PDE-based inverse problems. Bayesian methods are attractive for uncertainty quantification but assume knowledge of the likelihood model or data generation process. This assumption is difficult to justify in many inverse problems, where the specification of the data generation process is not obvious. We adopt a Gibbs posterior framework that directly posits a regularized variational problem on the space of probability distributions of the parameter.
We propose a novel model comparison framework that evaluates the optimality of a given loss based on its ``predictive performance''.
We provide cross-validation procedures to calibrate the regularization parameter of the variational objective and compare multiple loss functions. Some novel theoretical properties of Gibbs posteriors are also presented. We illustrate the utility of our framework via a simulated example, motivated by dispersion-based wave models used to characterize arterial vessels in ultrasound vibrometry.
\end{abstract}

\section{Introduction} \label{section1}

Quantification of uncertainty in the context of inverse problems is increasingly demanded by many applications \citep{stuart2010}. Bayesian statistics provides a useful viewpoint for this demand \citep{cdrs2009}. In a Bayesian framework, one prescribes a prior distribution summarizing relative uncertainties about possible solutions to the inverse problem. After observing noisy data, one updates the probabilities to obtain a posterior distribution of the possible solutions.
A fundamental component of a Bayesian formulation is the data-generating process or \emph{likelihood}. Specification of a likelihood is often invoked as a necessary condition to
guarantee theoretical properties of the posterior. However, it is difficult to specify the data-generating 
process in nonlinear inverse problems due to two main sources of model uncertainty:
forward model uncertainty, with respect to the underlying system dynamics; uncertainty, or lack of knowledge, with respect to the distribution of noise.
The possibility of \emph{model misspecification} raises a serious concern about using Bayesian methods. In this paper, we propose to solve inverse problems using an alternative, \emph{Gibbs posterior} or \emph{Generalized Bayes} framework, proposed by \cite{jt2008,bhw16,stability,zmaa}. The framework similarly requires a prior distribution and outputs a probability update conditional on the data. Gibbs posteriors do not rely on the knowledge of likelihood. They are derived as a solution to a variational problem on the space of probability measures on the space of solutions. They require the choice of a \emph{loss function} that measures the mismatch between the model and the data.

To use the Gibbs/generalized Bayes approach to solve inverse problems, several questions need to be addressed. Without knowledge of the underlying data-generating mechanism, how can we make a good choice of loss? In the variational objective that needs to be minimized, how do we determine the regularization parameter? The regularization parameter plays a vital role in balancing the trade-off between fidelity to the observed data and exploiting prior information. Finally, is the variational problem well-posed?

\subsection{Contributions} \label{section1_0}

The main contributions of this paper are the following:
\begin{enumerate}
    \item[1] A theory of model comparison for Gibbs posteriors that enables model comparison for loss functions. We define a notion of ``predictive performance'' for Gibbs posterior and study its theoretical properties.
    \item[2] We develop a particle filter and importance sampling method to simultaneously sample from the underlying Gibbs posterior and calibrate the regularization parameter that balances the loss function and regularization with respect to the prior. Our calibration procedure minimizes a novel leave-one-out cross-validation (LOOCV) objective. Due to the distributional nature of the solution, existing cross-validation algorithms are not immediately applicable.
    \item[3] We prove the stability and consistency of Gibbs posteriors. We show the continuity of Gibbs posterior as a mapping of the data in various distances for probability distributions. Our proposed upper bound improves on existing upper bounds that are vacuous when the perturbation to the data is large. We also study the asymptotic behavior of Gibbs posteriors in the large sample limit. The technical aspects of a consistency proof rely on tools in the robust Bayes estimation literature. We also study the asymptotics of a predictive distribution used for model selection associated with the Gibbs posterior.
\end{enumerate}
    
\subsection{Prior work} \label{section1_1}
The relation between the 
regularized least-squares problem proposed by \cite{tikhonov} and the maximum a posteriori (MAP) estimation problem in Bayesian statistics has been known for some time. Bayesian methods for inverse problems have been successfully adopted in diverse domains, nicely summarized by \cite{ks2005}. Recent literature \citep{cdrs2009,stuart2010,mcmc_function}
has extended the Bayesian framework with Gaussian likelihood to infinite-dimensional settings. The Gibbs posterior framework \citep{bhw16,jt2008,mmw2017} is not new, and its application in inverse problems was studied by \cite{zmaa,stability}. Similar concepts have been studied by \cite{gl2007,go17,md2018,bhattacharya2019bayesian}, among others, for improving the robustness of Bayesian inference under model misspecification. 
The novel model selection theory we develop in this paper can be viewed as an analog of the theory of Bayesian model selection and Bayesian cross-validation under model misspecification \citep{bayes}. Computationally, we rely on sequential Monte Carlo and particle filters algorithms. These algorithms have gained recent attention for potential use in Bayesian inverse problems. \citep{kantas2014sequential,beskos2015sequential} have used particle filters to solve parabolic and elliptic inverse problems. \cite{zmaa} have proposed a combination of particle filter and reduced order models for improved computational efficiency.

    A vast amount of literature exists on quantifying uncertainty in inverse problems. We place our method in context with previous ideas. Our variational formulation shares similarities with variational Bayes methods used in nonlinear inverse problems \citep{FRANCK2016215} and stochastic design \citep{KOUTSOURELAKIS2016124}. In these works, an objective involving a complex posterior distribution is minimized under the constraint that the approximating distribution is easy to sample from. In contrast, we use the variational problem to define the distribution of interest. 
 Second, when the likelihood is intractable, approximate Bayesian computation (ABC) has been proposed as a viable method of approximating intractable likelihoods \citep{lgass2015,zeng2019novel}.
However, these procedures can be computationally costly and do not address model misspecification. 

    Finally, several methods have been proposed for solving stochastic inverse problems and nonparametric probability measure estimation. Gradient-based optimization methods have been used to solve stochastic inverse problems \citep{narayanan2004stochastic, borggaard2015gradient,warner2015stochastic}. When the unknown parameter itself is a probability distribution, \cite{banks2015asymptotic,banks2015existence} have proposed minimizing a discretized objective stated in terms of the Prohorov metric. An exciting avenue we do not pursue in this work is to investigate possible connections between these non-Bayesian approaches and the Bayes/Gibbs posterior frameworks.

\subsection{Outline of the Paper} \label{section1_2}

Section \ref{section2} presents the foundations for the Gibbs posterior framework in the setup of inverse problems with model uncertainty. In Section \ref{section3}, we offer results on stability and asymptotic properties of Gibbs posteriors. We also present our novel contributions to model selection for different loss functions that are not intrinsically comparable to each other. In Section \ref{section4}, we describe a Monte Carlo algorithm that simultaneously learns the regularization parameter based on the LOOCV criterion and samples from the underlying Gibbs posterior. The algorithm is novel and relies on recent advances in particle filtering and importance sampling for Bayesian LOOCV. Section \ref{section5} presents numerical experiments illustrating the benefits of our approach. We conclude the paper with a discussion of future directions in Section \ref{section6}. All proofs are collected in the Appendix.

\section{Gibbs Posterior with Model Selection} \label{section2}

We review the foundations for the Gibbs posterior framework and describe properties of the Gibbs posterior {proposed by \cite{bhw16}}. Section \ref{section2_4} describes the problem of model comparison and our original contribution of predictive model selection theory for Gibbs posteriors.

\subsection{Notations} \label{section2_1}
We fix some notation used throughout the text. We denote as $||\cdot||$ the norm in an Euclidean space $\R^m$. We write as $\Delta(\mathcal{X})$ the space of all probability distributions on $\mathcal{X}\subset\R^m$, assuming standard Borel $\sigma$-algebras. For two probability measures $\mu,\nu\in\Delta(\mathcal{X})$, $d_{TV}(\mu,\nu),d_H(\mu,\nu),$ and $\kl(\mu,\nu)$ denote the total variation metric, Hellinger metric, and Kullback-Leibler (KL) divergence \citep{metrics}, respectively. For two probability measures $\mu,\nu$ (possibly on different spaces $\mathcal{X}_1$ and $\mathcal{X}_2$), we denote by $\mu\otimes\nu$ their product measure.
For a probability measure $\mu\in\Delta(\mathcal{X})$, $L^q(\mathcal{X};\mu)$ is the space of all functions $f:\mathcal{X}\to\R$ that are $L^q$-integrable with respect to $\mu$, where $q\in[1,\infty]$.

\subsection{Parametric Inverse Problems with Model Uncertainty} \label{section2_2}

Throughout this paper, we assume observing $n$ i.i.d. variables that take values in $\Y\subset\R^d$ with an unknown probability distribution $\P$:
\begin{equation} \label{iid}
    y_i \stackrel{iid}{\sim} \P\equiv\P_{\F(\theta_0)}.
\end{equation}
Here, the parameter $\theta_0$ is a physically meaningful parameter that characterizes the observed system. The parameter-to-observation map $\F(\theta)$ is often defined in relation to the parameterized PDE model: 
\begin{equation} \label{forward_model}
    \M(u(\theta);\theta) = 0,\; u(\theta)\in\mathcal{U},\;
    \M:\mathcal{U}\to\mathcal{V}^*.
\end{equation}
where $\mathcal{U},\mathcal{V}$ are Hilbert spaces with $\mathcal{V}^*$ being the dual space of $\mathcal{V}$. We assume for every $\theta$ there exists a unique $u(\theta)$ that satisfies \eqref{forward_model}. The parameter-to-observation map is defined as
$\F(\theta):=\mathcal{D} u(\theta)$, where $\mathcal{D}$ is the observation operator.

In the classical Bayesian framework, the parameterization of the sampling distribution \eqref{iid} by the forward model \eqref{forward_model} is \emph{known}. Examples include the additive white noise model \citep{kvzh2011} and Poisson likelihood \citep{barmherzig2022towards}. However, in practice, this need not be true because either the hypothesized parameterization is \emph{incorrect}, or the model uncertainties are so large that such a parameterization is difficult. 
Both errors in the forward model and errors in the noise distribution contribute to an incorrect parameterization of the likelihood. There are many ways in which such a mismatch can arise. A concrete example in ultrasound vibrometry application is reviewed in Section \ref{section5_2}. Here, we only mention that both philosophical and asymptotic justification of Bayesian inference is more tenuous with model misspecification. While the modeler may use a ``surrogate likelihood'' to define a misspecified Bayes posterior and argue that one can obtain good approximations when the surrogate is ``close'' to $\P_\theta$, defining this ``closeness'' in nonlinear inverse problems is not trivial.

In the next Section, we review a variational formulation that bypasses these difficulties. Instead of trying to define the correctly parameterized $\Q_\theta=\P_\theta$, the variational perspective \emph{defines} a discrepancy between the posited forward model and the observed data. The relative weights given to possible parameters $\theta$ are higher if they yield smaller discrepancy or loss.

\subsection{Variational Framework for Gibbs Posteriors} \label{section2_3}

Let $L:\Theta\times\R^d$ be a loss function. Let $\rho_0\in\Delta(\Theta)$.  We propose to solve the problem proposed by \cite{bhw16}:
\begin{equation} \label{gibbs_problem}
    \hrho_n^W(d\theta) := \argmin_{\rho \in \Delta(\Theta)} \left[ \bR_W(\rho) =  \int_\Theta \frac{1}{n}\sum_{i=1}^{n}
    L(\theta,y_i)~\rho(d\theta) + \frac{1}{nW}\kl(\rho||\rho_0) \right].
\end{equation}
Here, $\rho_0$ is the distribution quantifying our prior and $W>0$ is a regularization parameter that we assume is
given for now. If $\rho$ is not absolutely continuous with respect to $\rho_0$, the divergence is defined to be $+\infty$. Often we will abbreviate the average loss over all data by $R_n(\theta) := \frac{1}{n}\sum_{i=1}^{n} L(\theta,y_i)$. 

To ensure the existence of a solution, we will make assumptions on the structure of the problem, motivated by the assumptions of \cite{cdrs2009} and \cite{stuart2010}.
\begin{assump}
Let the loss $L(\theta,y)$ have the form $l(\F(\theta),y)$ and satisfy the following.
\begin{enumerate}
    \item $L(\theta,y)$ is uniformly bounded from below:
    \begin{equation*}
        \inf_{\theta, y} L(\theta, y) \geq B > -\infty.
    \end{equation*}
    We assume $B = 0$ without loss of generality. 
    \label{lower}
    \item For every $\theta,y$ there exists $K\equiv K(||\theta||,||y||)\in L^1(\Theta\times\Y;\rho_0\otimes\P)$ such that $L(\theta,y)\leq K(||\theta||,||y||)$.
    \label{upper}
    \item For every $r>0$ there exists $C_1(r,y)>0$ such that whenever $||\theta_1||,||\theta_2|| < r$,
    \begin{equation*}
        |L(\theta_1,y)-L(\theta_2,y)| \leq C_1(r,y)||\theta_1-\theta_2||
    \end{equation*}
    with $C_1(y)\equiv C_1(r,y)\in L^1(\Y;\P)$.
    \label{lipschitz_par}
    \item For every $r>0$ there exists $C_2(r,\theta)>0$ such that whenever $||y_1||,||y_2|| < r$,
    \begin{equation*}
        |L(\theta,y_1)-L(\theta,y_2)| \leq C_2(r,\theta)||y_1 - y_2||
    \end{equation*}
    with $\exp(C_2(\theta))\equiv C_2(r,\theta)\in L^1(\Theta;\rho_0)$.
    \label{lipschitz_data}
\end{enumerate}
\label{a_loss}
\end{assump}

\begin{rem}
    Note that because $L$ is defined to be a mapping on $\Theta\times\Y$, the regularity assumptions implicitly place restrictions on the forward model $\F$. We will use the squared $\ell^2$ loss as an example to understand the regularity conditions
    \begin{equation*}
        L(\theta,y) \equiv l(\F(\theta),y) = ||y - \F(\theta)||^2.
    \end{equation*}
    The properties of $\F$ dictate whether the loss $L$ satisfies the assumptions. Various PDE-based models used in the literature, combined with the popular Gaussian prior distribution, satisfy assumptions \ref{lower} and \ref{lipschitz_data} for squared $\ell^2$ loss; see, e.g., Section 3, \cite{stuart2010}. 
    On the other hand, the integrability conditions \ref{upper} and \ref{lipschitz_par} depend on the unknown $\P$. One can check how mild or severe these conditions turn are for a specific loss function, by hypothesizing models like \eqref{add_model}, without specifying a likelihood.
\end{rem}

We will also make a mild smoothness assumption on the density of the prior distribution $\rho_0$. The Gaussian prior satisfies the assumption.

\begin{assump}
$\rho_0$ has positive density everywhere. Furthermore, for every $r>0$ there exists $C_3(r)>0$ such that whenever $||\theta_1||,||\theta_2|| < r$,
    \begin{equation*}
        |\log\rho_0(\theta_1) - \log\rho_0(\theta_2)| \leq C_3(r)||\theta_1 - \theta_2||.
    \end{equation*}
    \label{a_prior}
\end{assump}

There exists a unique solution to \eqref{gibbs_problem} in $\Delta(\Theta)$, which has the following density for fixed $W>0$:
\begin{equation} \label{gibbs_formula}
    \hrho_n^W(d\theta) :=
    \frac{\exp\{-nWR_n(\theta)\}
    \rho_0(d\theta)}{Z_n^W},
\end{equation}
where the normalizing constant, or the ``partition function'' $Z_n^W$, is defined as
\begin{equation} \label{partition}
    Z_n^W\equiv\int e^{-nWR(\theta)}~\rho_0(d\theta).
\end{equation}
To derive this formula, the objective functional can be rewritten as
\begin{equation} \label{decomposition}
    \bR_W(\rho) = \frac{1}{nW}\left\{\kl(\rho||\hrho_n^W) - \log Z_n^W\right\};
\end{equation}
The first term is non-negative and uniquely attains zero at $\rho\equiv\hrho$. The second term does not depend on $\rho$, so the minimum of the functional is achieved at $\bR_W(\hrho_n^W) = -\log Z_n^W$. The possible technical issues are measurability of $\exp(-nWR_n(\theta))$ and finiteness of $\log Z_n^W$, which follow from Assumption \ref{a_loss}.

\begin{rem}
When we fix $W=1$ and choose the loss
to be the negative log-likelihood: $L(\theta,y)=-\log p(y|\F(\theta))$, 
the Gibbs posterior coincides with a Bayes posterior update, using $p(y|\F(\theta))$ as its likelihood component.
Thus, our Gibbs posterior solution strictly generalizes the Bayes posterior.
\end{rem}

We close the Section with some intuition of the role of $W$ in the Gibbs posterior. In the limit $W\to 0$, $\hrho_n^W\equiv\rho_0$, so there is no update of information from the prior. Smaller $W$ thus more heavily weighs prior information. In the limit $W\to\infty$, $\hrho_n^W$ concentrates on a set of $\theta$s minimizing the loss over the observed data. Larger $W$ thus more heavily weighs information from the data and leads to increased sensitivity to perturbation under noise.
Intuition suggests that a prior $\rho_0$ that is strictly positive on $\Theta$ should reflect large uncertainty and that $W$ must be carefully chosen based on the amount of information in the data relative to the prior. Inspection of \eqref{gibbs_problem} suggests that we are implicitly implementing a discrepancy principle \citep{nair2009morozov} since the divergence penalty has less influence when the sample size is large.

\subsection{Extension to Model Selection} \label{section2_4}

\subsubsection{Predictive Model Selection}
\label{section2_4_1}

Solving the variational problem \eqref{gibbs_problem} still requires a pre-specified choice of loss $L$. It may appear this requirement is as restrictive as positing the generating process as a better choice of loss
 hinges on knowledge or assumptions on $\P$. Our first proposal is to \emph{define} a valid way to compare two different losses without
 requiring knowledge of the data-generating mechanism. The key idea is to compare them based on the ability to make accurate \emph{predictions}, measuring their discrepancy on a future observation. As mentioned, this principle is not new and has been used to improve the robustness of Bayesian model prediction and model checking. The novelty lies in the definition of the predictive density without assuming the likelihood, which will serve as a natural discrepancy measure between the new observation and prediction.

Consider a common prior distribution $\rho_0$ and multiple competing losses, $L_1,\ldots,L_k$, defined on subsets $\Theta_1,\ldots,\Theta_k$ of $\Theta$. Given the corresponding set of Gibbs posteriors $\hrho_{n,1}^{W_1},\ldots,\hrho_{n,k}^{W_k}$, we propose the following predictive model comparison principle:
map each distribution $\hrho_{n,m}^{W_m}$ to a  prediction taking values in $\Y$ and measure its discrepancy from another measurement on the observation space, $\ynew$. The solution minimizing this prediction discrepancy, i.e.
\begin{equation*}
    \argmin_{m\in\{1,\ldots,k\}} D_{pred}(\ynew;\hrho_{n,m}^{W_m}),
\end{equation*}
for a common discrepancy metric $D_{pred}$, is chosen to be optimal.

\subsubsection{Gibbs Predictives}
\label{section2_4_2}

To implement the predictive model selection strategy, we need to choose the predictive discrepancy measure $D_{pred}$. In Bayesian statistics, similar problems arise when considering $k$ different competing likelihood models describing the data-generating mechanism, none of them required to be correct. The default choice is the log-score of the \emph{posterior predictive distribution} \citep{vgg17}:
\begin{equation} \label{ppd}
    -\log p_{n,m}(\ynew) := -\log\int p_m(\ynew|\theta)~p_{n,m}(\theta)d\theta,
\end{equation}
where $p_{n,m}(\theta)$ is the Bayes posterior for the $m$-th model after observing $n$ data points, and $p_m(y|\theta)$ defines the model likelihood. The formed posterior predictive makes a distributional prediction, similar to how the posterior makes a probabilistic estimation by giving different probabilities to possible parameter values. The likelihood model that places the highest posterior predictive density on the next measurement made, $\ynew$, is chosen to be an optimal description.

The second proposal in this Section is to generalize the notion of model selection based on posterior predictive score \eqref{ppd}, similar to how we extended the notion of Bayes posterior in Section \ref{section2_3}. This is a non-trivial problem because evaluating predictive power based on \eqref{ppd} already assumes a hypothesized likelihood. Our general Gibbs posterior framework is more flexible, as it allows a general form of losses, but then we also cannot make sense of a natural distributional prediction \eqref{ppd}. Nevertheless, we define below the notion of a Gibbs predictive distribution that subsumes the case of Bayes predictive distributions.

\begin{definition}
    Let $\lambda$ be a Lebesgue density of a probability measure on $\Y$. We define the Gibbs predictive distribution of $\ynew$ given $\y\equiv(y_1,\ldots,y_n)$ by its Lebesgue density, with respect to $\lambda$:
    \begin{equation} \label{predictive}
        \hat{p}_n^W(\ynew) := \frac{\int\exp\{-L(\theta,\ynew)\}~\hrho_n^W(d\theta)\lambda(\ynew)}
        {\iint\exp\{-L(\theta,\ynew)\}~d\lambda(\ynew)~
        \hrho_n^W(d\theta)}.
    \end{equation}
    \label{gibbs_pd}
\end{definition}

\begin{rem}
Definition \ref{gibbs_pd} subsumes the Bayes posterior predictive distributions as a special case. We choose the loss to be $-\log p(y|\theta)$ for some posited likelihood, restrict $W=1$, and set $\lambda(y')\equiv 1$, to obtain:
    \begin{equation*}
        \hat{p}_n^1(\ynew) = \frac{\int p(\ynew|\theta)~p(\theta|\y)d\theta}{\iint p(y'|\theta)~p(\theta|\y)d\theta~dy'} \equiv p(\ynew|\y),
    \end{equation*}
    since the denominator is automatically 1. The formal similarity suggests that we are proposing to define a predictive distribution based on the ``likelihood'' $\exp(-L)\times\lambda$. This generalization is only formal when $\Y$ is non-compact because $\lambda\equiv 1$ is no longer a probability measure. The denominator of \eqref{predictive} is often still finite, as is the case for all loss functions used in the numerical experiments in Section \ref{section5} below, but that does not immediately follow from the regularity assumptions \ref{a_loss}.
\end{rem}

Definition \ref{gibbs_pd} completes the specification of our novel model selection principle in inverse problems under model uncertainty. Given different Gibbs posterior solutions $\hrho_{n,1}^W,\ldots,\hrho_{n,k}^W$ derived under a common prior $\rho_0$ and different losses $L_1,\ldots,L_k$, we propose to choose the optimal solution by the criterion
\begin{equation*}
    \argmin_{m\in\{1,\ldots,k\}} -\log\hat{p}_{n,m}^{W_m}(\ynew).
\end{equation*}
Given a set of $n$ measurements $y_1,\ldots,y_n$ we implement an approximation of the criterion using LOOCV:
\begin{equation} \label{pred_criterion}
    \argmin_{m\in\{1,\ldots,k\}} \left\{
    P_{CV}(m) := -\frac{1}{n}\sum_{i=1}^{n}
    \log\hat{p}_{(-i),m}^{W_m}(y_i)
    \right\},
\end{equation}
where $\hat{p}_{(-i),m}^{W_m}$ indicates the $m$-th Gibbs posterior solution derived from the dataset minus a datum $y_i$. Computational aspects will be discussed in the next Section, together with calibrating $W_m$, for which we again resort to LOOCV. The asymptotics for predictions using our construction as $n\to\infty$ are discussed in Section \ref{section4}.

We conclude this discussion with a remark on computing pointwise evaluations of \eqref{gibbs_pd} involving high-dimensional integral $\tilde{Z}(\theta)$ nested inside an integral over $\Theta$. Losses used in practical inverse problems often have enough local structure to yield an integral that can be decomposed into low-dimensional subparts. Furthermore, some popular but simple losses like squared $\ell^2$-norm loss lead to analytically tractable integrals. A recipe for evaluating $\tilde{Z}(\theta)$ is only possible by utilizing a more detailed structure of the problem. We will show some examples in Section \ref{section5}.

\section{Model Calibration and Computation} \label{section3}

In almost all practical scenarios, probability distributions of the form \eqref{gibbs_formula} do not admit a closed-form density, so we focus on sampling algorithms for Gibbs posteriors. 
MCMC algorithms allow us to draw samples that can target posteriors \citep{bda3}. As long as the regularization parameter $W$ is given, the computational problem in our framework is equivalent to Bayesian methods. The added complexity is to find $W$ given the data.

We mention two important precursors that can be used to learn $W$. The first is the SafeBayes algorithm \citep{go17}. In view of theoretical results by \cite{gm20}, mentioned in the previous Section, it attempts to construct a Gibbs posterior that mimics the performance of that for an oracle choice of $W$. The other is the generalized posterior calibration algorithm \citep{sm19}, which chooses $W$ so that the Gibbs posterior intervals satisfy valid frequentist coverage.

In inverse problems, one often deals with small sample size $n$ and a forward model $\F$ is expensive to repeatedly evaluate. Calibrating frequentist coverage can be problematic conceptually (due to small $n$) and computationally (reliance on bootstrap methods). To address these concerns, when $n$ is moderately large, we discount the value of $W$ based on the expected loss estimated through cross-validation. This step is quite similar to the SafeBayes algorithm, but less computationally demanding. Our method can be seamlessly integrated into the particle filter sampling algorithm for a sequence of Gibbs posteriors. The model selection procedure presented in Section \ref{section2_4} also relies on a cross-validation objective \eqref{pred_criterion} and can be implemented using a particle filter.

\subsection{Cross-Validation with Multiple Samples} 
\label{section3_1}

Our choice of sampling method is motivated by the need to calibrate the regularization parameter $W$ based on the data.
When $n > 1$ sample observations of the underlying function are observed, frequentist statistics can be invoked to make a conscientious choice of $W$. The optimal choice of $W$ is defined to minimize the error our inverse solution makes outside the observed data:
\begin{align}
    R(W) = \E\left[\int R_n(\theta)~d\hrho_n^W(\theta)\right],
    \label{expected_risk}
\end{align}
where the expectation is with respect to the observed data $y_1,\ldots,y_n$.
\eqref{expected_risk} is a natural risk to minimize from a frequentist perspective, but we cannot compute it. Instead, we can estimate it through cross-validation. In the special case when each test set consists of a single $y_i$, we obtain the leave-one-out cross-validation (LOOCV) estimate of $R(\alpha)$ \citep{eslii}:
\begin{align}
     R_{CV}(W) = \frac{1}{n}\sum_{i=1}^{n}\int L(\theta,y_i)~d\loo{\hrho}^W(\theta).
     \label{obj_cv}
\end{align}
Here, $\loo{\hrho}^W$ denotes a Gibbs posterior derived as in \eqref{gibbs_formula}, except that we hold out $y_i$ from the dataset $\y$.

{LOOCV is a popular choice for choosing regularization parameters in linear inverse problems since \cite{gcv}, and there are many variants of the GCV algorithm. However, they are not applicable to our setting because they do not address the fundamental difficulty of computing expectations under different probability distributions with changing $W$. 
Instead, we turn to sampling. We set a grid of $\{W_0,W_1,\ldots,W_T\}$, fixing $W_0\equiv 0$ and $W_1\equiv\bar{W}$ for an upper bound $\bar{W}$. The goal is to approximate all expectations involved in \eqref{obj_cv} by Monte Carlo samples drawn from the corresponding distributions. A naive Monte Carlo approach is expensive as it requires sampling from a sequence of in total $nT$ different probability distributions.
Thus, we need a sampling method that uses a sequence of distributions that can approximate all expectations involved in \eqref{obj_cv}. In this section, we will show how importance sampling based on a carefully chosen sequence of $T$ distributions allows us to estimate \eqref{obj_cv}.}

\subsection{Importance Sampling Cross-Validation} \label{section3_2}

Importance sampling yields a useful approximation when considering mild changes in the posterior distribution \citep{bda3}. {We first determine a ``proposal distribution'' $\tilde{\rho}\equiv\tilde{\rho}^W$ for each value of $W$.} Then, the formula for estimating $R_{CV}$ using $S$ Monte Carlo draws from $\tilde{\rho}$ is
\begin{align}
    \hat{R}_{CV}(W) = \frac{1}{n}
    \sum_{i=1}^{n}
    \frac{\sum_{s=1}^{S} r_i^W(\theta^{(s)})L(\theta^{(s)},y_i)}
    {\sum_{s=1}^{S}r_i^W(\theta^{(s)})},
    \label{is_estimator}
\end{align}
where the importance weights have the formula
\begin{align}
    {r_i^W(\theta) = \frac{\exp\{-W\sum_{j:j\neq i}L(\theta,y_j)\}}{\tilde{\rho}(\theta)}.}
    \label{loo_weight}
\end{align}
{The choice of $\hrho$ is crucial to numerical stability of \eqref{is_estimator}. For example, a seemingly natural choice of $\tilde{\rho}$ is the ``full'' posterior, $\hrho_n^W$, but it has a known problem of importance weights having possibly} high or infinite variance \citep{vgg17,cd18}.
{Recently, \cite{silva2022robust} have proposed a different choice of $\rho^t$ that produces importance weights with finite asymptotic variance. The new distribution for each $t$ has a density representation
\begin{equation} \label{mix_instr}
    \rho_{mix}^W(\theta) := 
    \frac{\sum_{i=1}^{n}\exp\{-W\sum_{j:j\neq i} L(\theta,y_j)\}}{\sum_{i=1}^{n}\int\exp\{-W\sum_{j:j\neq i} L(\tilde{\theta},y_j)\}~d\rho_0(\tilde{\theta})}.
\end{equation}}
{We set our choice of $\tilde{\rho}$ for each $W$ to be precisely \eqref{mix_instr}. Computation of importance weights \eqref{loo_weight} requires tractability of the numerator of density \eqref{mix_instr}, as estimator \eqref{is_estimator} renormalizes the weights. As long as we can draw samples from each $\rho_{mix}^W$ for $W\in\{W_1,\ldots,W_T\}$, we can estimate \eqref{is_estimator}. This reduces the burden of sampling from targeting $nT$ distributions to just $T$ distributions.}

{We can write explicit importance sampling estimators for both calibration of $W$ and model selection \eqref{pred_criterion}. Let $\theta^{(1)},\ldots,\theta^{(S)}$ be (approximately) drawn from distribution \eqref{mix_instr}.
The calibration objective \eqref{obj_cv} is approximated by
\begin{equation} \label{approx_obj_cv}
    \frac{1}{n}\sum_{i=1}^{n}
    \frac{\sum_{s=1}^{S}r_i^W(\theta^{(s)})
    L(\theta^{(s)},y_i)}
    {\sum_{s=1}^{S}r_i^W(\theta^{(s)})}
\end{equation}
The model selection objective \eqref{pred_criterion} is approximated by
\begin{equation} \label{approx_pred_crt}
    -\frac{1}{n}\sum_{i=1}^{n}\log\left(
\frac{\sum_{s=1}^{S}r_i^W(\theta^{(s)})e^{-L(\theta^{(s)},y_i)}}{\sum_{s=1}^{S}r_i^W(\theta^{(s)})\tilde{Z}(\theta^{(s)})}
    \right) = 
    \frac{1}{n}\sum_{i=1}^{n}\left(\log\sum_{s=1}^{S}r_i^W(\theta^{(s)})\tilde{Z}(\theta^{(s)}) - 
    \log\sum_{s=1}^{S}e^{(W-1)L(\theta^{(s)},y_i)}\right).
\end{equation}
}

\subsection{Particle Filter Approximation}
\label{section3_3}

Usually, it is not possible to draw i.i.d. samples from any one distribution of {the sequence $\rho_{mix}^{W_t}$ for increasing $t$.} Particle filtering, or sequential Monte Carlo (SMC) algorithms, comprise a suite of techniques to obtain a sequence of $S$-sized particle approximations for each distribution.
While filtering methods are often discussed in the context of estimating time-varying parameters, they are easily adapted to our setting by equating ``time'' with $t$.

We start with i.i.d. draws $\{\theta_0^{(1)},\ldots,\theta_0^{(S)}\}$ drawn from the prior, $\hrho_0\equiv\rho_0$, {since $W_0=0$}. At each iteration $t$, an SMC sampler implements three moves:
\begin{enumerate}[label=(\alph*)]
    \item (Weighting) Each sample is weighted according to the importance sampling formula
    \begin{equation*}
{        
    w_t(\theta_{t-1}^{(s)}) \propto
    \frac{\rho_{mix}^t(\theta_{t-1}^{(s)})}{\rho_{mix}^{t-1}(\theta_{t-1}^{(s)})}
}
    \end{equation*}
    where $\rho_{mix}^t\equiv\rho_{mix}^{W_t}$. {Since the normalizing constants are unknown, each weight is renormalized to \begin{equation*}
        \bar{w}_t(\theta_{t-1}^{(s)}) := \frac{w_t(\theta_{t-1}^{(s)})}{\sum_l w_t(\theta_{t-1}^{(l)})}.
    \end{equation*}}
    \label{is_step}
    \item (Resampling) $S$ new particles $\{\tilde{\theta}^{(s)}_{t-1}\}$ are drawn with probabilities 
    \begin{equation*}
        Pr(\tilde{\theta}^{(s)}_{t-1} =  \theta_{t-1}^{(s')}) = \tilde{w}(\theta_{t-1}^{(s')}),
    \end{equation*}
    for possible pairings $(s,s')$.
    \label{res_step}
    \item (Mutation) $S$ Markov chains are run in parallel for $K>0$ steps, each having $\bar{\theta}_{t-1}^{(s)}$ as initial states. The transition kernel of each chain is constructed to have an invariant measure $\rho_{mix}^t$. $\theta_t^{(s)}$ is assigned the value of the draw from a Markov chain run for $K$ iterations.
    \label{mcmc_step}
\end{enumerate}
The mutation step can be implemented using standard MCMC methods, such as Metropolis-Hastings with a Gaussian proposal. All steps in this algorithm are parallelizable across $S$ particles, an attractive feature for computational efficiency.

\subsection{{Practical Considerations}} 
\label{section3_4}

{The sampling method described has several  hyperparameters that are potentially critical to reliable calibration and model selection.}

{First is the choice of grid size. For an upper bound, we fix $\bar{W} \equiv 1$ by normalizing the loss $L$ with an appropriate scale, either physically meaningful or learned from the data. A number of standard estimators like the variance of the data can be used. We can even learn $W$ with uncertainty, extending the loss function to include $W$ as an unknown parameter \citep{bhw16}. 
The grid spacing is also important. \cite{go17} suggest a grid of $(0,1)$ that are spaced with exponentially growing width. This is at odds with the computational stability of filtering methods, and indeed \cite{smc} suggest that a grid must be spaced with at least a geometrically decreasing width. 
Their theoretical results and suggestion do not exactly apply to the properties of the particular sequence of distributions we consider. In numerical implementations, we used the growing-width sequence of \cite{go17}. Our observation is that \eqref{ess} can quickly drop at the initial move from $W_0=0$ to $W_1$, but remain relatively stable throughout the rest of the sequence. If importance weights degenerate quickly, further adaptive tempering steps by \cite{jasra11} may be incorporated to interpolate between two steps $W_t$ and $W_{t+1}$.}

{Second, we must choose the number of samples $S$ and the number of Markov chain transitions $K$ in step \ref{mcmc_step}. Ideally, these should not be tuning parameters but chosen based on theoretical bounds on the numerical approximation errors induced by approximating $\hrho_n^t$ with a discrete measure based on the particles. However, the current state-of-the-art bounds appear not sharp enough for many practical purposes. In general, increasing $S$ and $K$ trades better approximation with higher computational costs. Practically, one must tune the parameters based on computational budget and some experience.}

{Third, we must observe steps \ref{is_step} and \ref{mcmc_step} require computing the loss at a new value of the forward map $\F(\theta)$, defined as a solution of a PDE. The computational cost for solving a PDE can ramp up quickly, as the mutation step requires $K$ repeated evaluations of the PDE solver for $S$ particles. We follow \cite{smc} and trigger steps \ref{res_step} and \ref{mcmc_step} only when importance weights show signs of degeneracy by diagnosing it with the effective sample size statistic (ESS):
\begin{equation} \label{ess}
    \mathrm{ESS}^t = \frac{(\sum_{s=1}^{S} w_t(\theta_{t-1}^{(s)}))^2}{\sum_{s=1}^{S} w_t(\theta_{t-1}^{(s)})^2}.
\end{equation}
ESS is bounded between 0 and 1, higher when the distribution of the importance weights is wider; a necessary but not sufficient condition for good approximation guarantees of importance sampling-based estimate. The algorithm will only trigger steps \ref{res_step} and \ref{mcmc_step} if $ESS^t$ falls below a prescribed threshold.}

{Finally, we note that for applications in which few measurements are assumed independent (small $n$), greedy minimization of \eqref{obj_cv} is unwise, given the statistical noise inherent in small samples. We follow the suggestion of ``one standard error rule'' \citep{eslii}, which chooses the smallest $W_t$ whose associated value of $R_{CV}$ is one standard error within the value of $R_{CV}$ at the minimizer in the grid. This approach favors smaller values of $W$ than exactly minimizing \eqref{obj_cv}, assuming it is better to be conservative about uncertainties when the model error is potentially large.}

\section{{Theoretical Analysis}}
\label{section4}

In this Section, we present {various desirable theoretical properties of} \eqref{gibbs_formula}. First, extant stability results for Bayesian inverse problem solutions \citep{cdrs2009,stuart2010} {are improved to capture the correct order of growth for the metric used to measure perturbation in the posteriors. Second,} we review statistical results that ensure consistency of Gibbs posteriors as $n\to\infty$. Our goal is not to formulate an all-encompassing theory of {consistency, as this warrants a whole work with attention to the detailed structure of specific inverse problems}. Rather, we aim to provide basic theoretical justifications of our framework {by showing both the Gibbs posterior on the parameter space and the Gibbs predictive on the observable space exhibit reasonable asymptotic convergences.}

\subsection{Continuity in Data} \label{section4_1}

The first stability result is on the continuity of Gibbs posterior in the underlying data. {There exist many metrics for probability distributions, so one can obtain stronger theoretical guarantees depending on the choice. \cite{cdrs2009} and \cite{stuart2010} have first presented stability results for \emph{Bayes posterior} that bound the \emph{Hellinger distance} from above by the distance between the underlying data. Since Hellinger distance is bounded above by 1, their bound is vacuous when the perturbation in the data is large. Below, we present a tighter, non-vacuous upper bound that is always at most 1. Our result first bounds the KL divergence between posteriors and exploit inequalities between different metrics of probability measures due to \cite{bretagnolle1979estimation}. Although KL divergence is asymmetric, our bound is valid for ``both directions.''}

\begin{theorem}
Let $r>0$, $W>0$, and $\y_1,\y_2\in\Y^n$ satisfying $\max_{i=1}^n \{||y_{1,i}||,||y_{2,i}||\}<r$. For two corresponding posteriors $\hrho_{n,1}^W$ and $\hrho_{n,2}^W$, there constants $a,b$ that only depend on the prior $\rho_0$ and $r>0$, such that
\begin{equation*}
{
\kl(\hrho_{n,1}^W||\hrho_{n,2}^W)\vee\kl(\hrho_{n,2}^W||\hrho_{n,1}^W) \leq 
    aW(1 + e^{bnW})^2\sum_{i=1}^{n}||y_{1,i} - y_{2,i}||^2.    
}
\end{equation*}
{As a consequence,
\begin{equation*}
    d_H^2(\hrho_{n,1}^W,\hrho_{n,2}^W)\leq \sqrt{1 - \exp\left\{-aW(1+e^{bnW})^2\sum_{i=1}^{n}||y_{1,i} - y_{2,i}||\right\}}.
\end{equation*}
}
\label{thm_stability}
\end{theorem}

The Gibbs posterior is also continuous in the regularization parameter $W$. The continuity argument justifies learning $W$ from the data as in Section \ref{section4}. The result is an easy consequence of the proof of Theorem \ref{thm_stability} and thus stated without proof.

\begin{theorem}
Suppose $y\in\Y$. Let $W,W'>0$. Then the map $W\mapsto\hrho_n^W$ is continuous in the Hellinger distance.
\label{continuity_W}
\end{theorem}

\subsection{Finite Approximation} \label{section4_2}

In practice, the exact PDE model \eqref{forward_model} is approximated using discretization schemes like finite elements methods. In this case, \eqref{forward_model} is replaced with a discretized counterpart:
\begin{equation*}
    \M^h(u^h(\theta);\theta) = 0,\; u^h\in\mathcal{U}^h,
\end{equation*}
where $\mathcal{U}^h$ is a finite dimensional projection of $\mathcal{U}$. One then considers a sequence of discretization level indices $h$ that decays to zero. Assuming each discrete forward model is well-posed, we can evaluate an approximate forward map $\F^h:\Theta\to\R^d$.
When $u^h(\theta)$ converges to $u(\theta)$ for each $\theta$, we desire convergence of the Gibbs posterior estimated using $\F^h(\theta)$ to that estimated using $\F(\theta)$. This is our next result. Again, a weaker result has been derived by \cite{stuart2010} for Bayesian problems, {and the bounds are improved similarly as in Theorem \ref{thm_stability}.}

\begin{theorem}
Define a surrogate loss $L^h(\theta,y) = l(\F^h(\theta),y)$, and suppose the new loss satisfies Assumption \ref{a_loss}. Furthermore, suppose for every $r>0$ and for every $h$, there exist $C(\theta)\equiv C(r,\theta)$ such that whenever $||y||<r$,
\begin{equation*}
    |L(\theta,y) - L^h(\theta,y)| = C(\theta)\psi(h),\;
    C(\theta)\in L^1(\Theta;\rho_0),
\end{equation*}
where $\psi(h)\to0$ as $h\to0$. Then, for two corresponding Gibbs posteriors
$\hrho_{n}^W$ and $\hrho_{n,h}^W$, there exist constant $a,b$ that only depend on the prior $\rho_0$ and $r>0$ such that
\begin{equation*}
{
    \kl(\hrho_n^W||\hrho_{n,h}^W)\vee
    \kl(\hrho_{n,h}^W||\hrho_n^W) \leq 
    aW(1 + e^{bnW})^2\psi(h).    
}
\end{equation*}
{As a consequence,
\begin{equation*}
    d_H^2(\hrho_n^W,\hrho_{n,h}^W)\leq
    \sqrt{1 - \exp\left\{
    -aW(1+e^{bnW})^2 \psi(h)
    \right\}}.
\end{equation*}
}
\label{thm_approx}
\end{theorem}

\subsection{Statistical Consistency} \label{section4_3}

The preceding stability results have been derived conditional on the data $\y$. Frequentist analysis of a sequence of
Gibbs posteriors $(\hrho_n^W)_{n=1}^\infty$, for a sequence of growing length, reveals the asymptotic convergence properties of our solution. To rigorously define consistency, we need to appropriately define convergence
of a sequence of posteriors $(\hrho_W^n)_{n=1}^\infty$. {In Bayesian statistics, a sequence of posterior distributions $\rho_n$ is said to be \emph{consistent} if
\begin{equation} \label{consistency1}
    \rho_n\to\delta_{\theta_0}\quad\text{with probability 1};
\end{equation}
i.e., the distribution converges to a Dirac measure at a true, unknown model parameter $\theta_0$, in the weak topology of probability measures.}
In typical consistency proofs, {we use another definition that is equivalent if $\Theta$ is a metric space \citep{gv2017}:}
\begin{equation} \label{consistency2}
    \rho_n(\Theta\setminus U)\to 0\quad
    \text{with probability 1},
\end{equation}
{for every open neighborhood $U$ of $\theta_0$.}
Recall now the general Gibbs posterior framework accounts for the possibility of model misspecification. {While it is possible there exists a physically meaningful true parameter $\theta_0$, the true likelihood parameterization $\P_{\theta_0}$ is unknown. Therefore, the same definition of ``convergence to truth'' is inappropriate.} 

It is reasonable to instead consider convergence to a set of parameters that minimize
the following expected loss \citep{kleijn2012bernstein}:
$$
\Theta^* := \left\{\theta^*:R(\theta^*) = \min_\theta R(\theta),\; R(\theta) = \int L(\theta,y)~\P(y)dy \right\},
$$
{It is easily checked that by Assumption \ref{a_loss}, $R$ is continuous in $\theta$ (c.f. proof in appendix). Therefore, if $\Theta$ is compact, $\Theta^*$ is guaranteed to be non-empty. We also note that in general it is not guaranteed that $\theta_0\in\Theta^*$.}

{Consistency in this context is defined as
\begin{equation*}
    \hrho_n^W(\Theta\setminus U)\to 0\quad\text{with probability 1,}
\end{equation*}
for every open set $U\supset\Theta^*$, generalizing \eqref{consistency2}.
The following theorem, essentially a consequence of theorems due to \cite{kleijn2012bernstein}, provides sufficient conditions under which \eqref{consistency2} describes the asymptotic behavior of Gibbs posteriors.}

\begin{theorem}
In addition the assumptions in Section \ref{section2}, suppose furthermore
\begin{enumerate}
    \item {$\Theta$ is compact;}
    \label{compact}
    \item {There exists $\bar{W}$ such that $C_1(y)e^{WL(\theta^*,y)}\in L^1(\Y;\P)$ for $C_1(y)$ in Assumption \ref{a_loss}.\ref{lipschitz_par}.}
    \label{tail}
\end{enumerate}
{
Then, $\Theta^*$ is non-empty. Furthermore, for every $W\in (0,\bar{W})$ and every open set $U\supset\Theta^*$,
\begin{equation*}
    \hrho_n^W(\Theta\setminus U)\to 0\quad\text{with probability 1.}
\end{equation*}
}
\label{thm_consistency}
\end{theorem}

\begin{rem}
    {
    Condition \ref{tail} does not follow from previously stated regularity assumptions.
    That $W$ must be small enough is intuitively reasonable, as the posterior becomes less stable with large $W$ at any finite $n$. For the technical subtleties of this condition, see \cite{sm23}. 
    We point out that mere continuity of $L$ and compactness of $\Theta$ are in general insufficient to guarantee consistency results under model misspecification as studied by \cite{kleijn2012bernstein}.
    }
\end{rem}

{In Section \ref{section2_4} and \ref{section3}, we proposed that an optimal loss must be chosen based on its predictive performance. Asymptotic analysis of the Gibbs predictive, \eqref{gibbs_pd}, can provide one justification of why our procedure is reasonable. The following theorem shows that if given \ref{thm_consistency}, the corresponding sequence of Gibbs predictives weakly converges to a certain ``likelihood'' involving the recovered $\theta*$. An interpretation is that even though we did not explicitly specify a generative model, our predictive criterion \eqref{pred_criterion} asymptotically learns a likelihood model that uses the best possible parameter estimate.}

\begin{theorem}
{
    Suppose $\Theta^* = \{\theta^*\}$. Assume further conditions given by Theorm \ref{thm_consistency}, so that $\hrho_n^W$ is consistent. Then, for a sequence of Gibbs predictives $\hat{p}_n^W(\ynew)$,
    \begin{equation*}
        \kl(\P||\hat{p}_n^W) - \kl(\P||\tilde{p}_{\theta^*})\to 0
        \quad\text{with probability 1,}
    \end{equation*}
    where $\tilde{p}_{\theta^*}$ has the density
    \begin{equation*}
        \tilde{p}_{\theta^*}(\ynew) = \frac{\exp\{-L(\theta^*;\ynew)\}\lambda(\ynew)}
        {\int \exp\{-L(\theta^*;\ynew)\}d\lambda(\ynew)}.
    \end{equation*}
}
    \label{thm_pred_const}
\end{theorem}

\section{Numerical Illustrations} \label{section5}

\subsection{Toy Example} \label{section5_1}

\begin{figure}[t]
    \centering
    \includegraphics[width=7in]{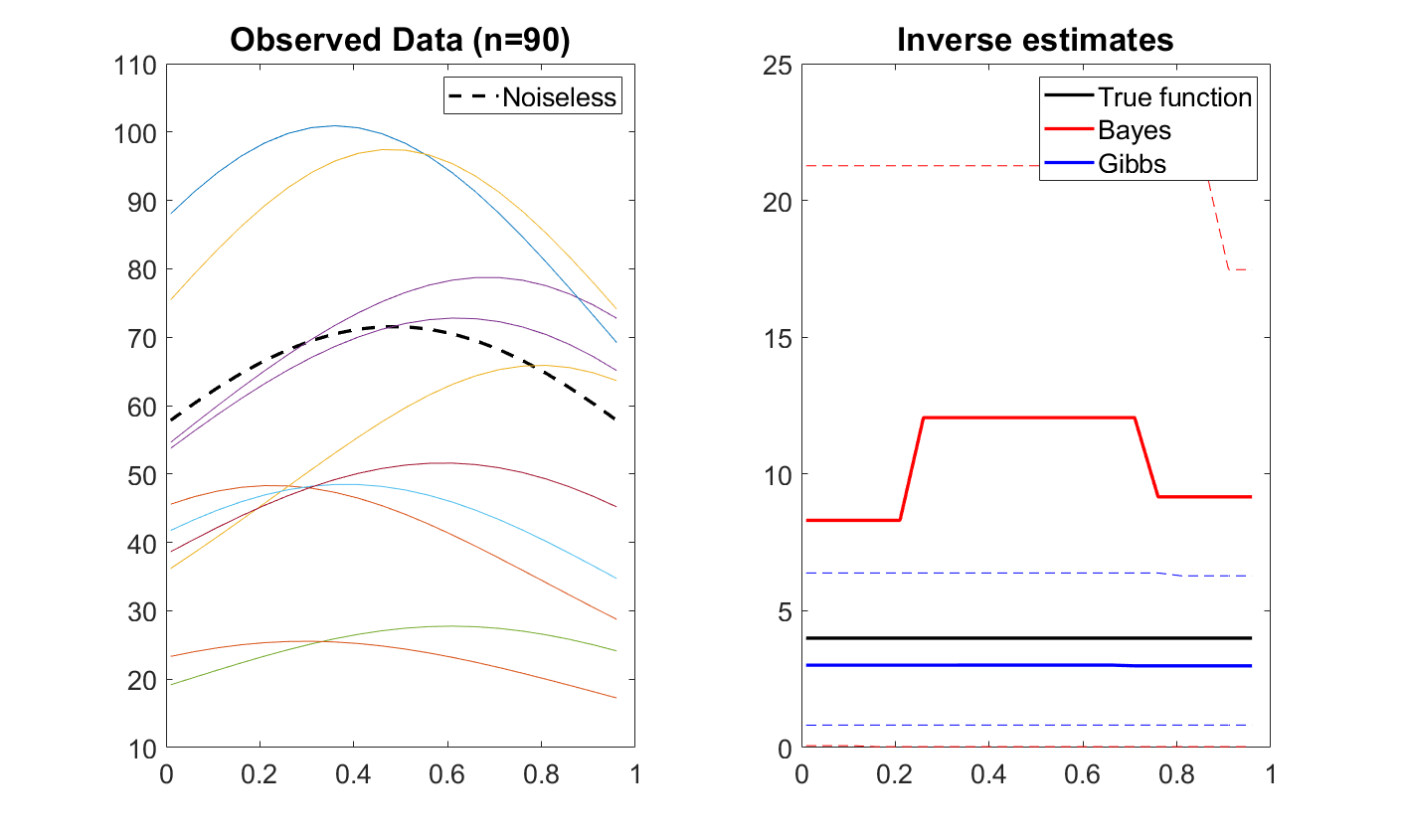}
    \caption{(Left) Simulated observed data for the experiment in Section \ref{section5_1} ($n=90$). Only ten subsamples are plotted for visual purposes.
    (Right) Comparison of Bayes and Gibbs posterior estimates for $u(\theta)$. Both mean estimates and 90\% quantile-based credible regions (dashed) are plotted.}
    \label{fig_toy}
\end{figure}

In this Section, we study a toy problem with an antagonistic data-generating process to highlight the advantage of uncertainty quantification outside a strictly Bayesian framework. {Consider a linear inverse problem in which the true solution, unknown to the researcher, is simply a constant.}  We consider estimation based on a family of piecewise constant functions indexed by a parameter vector $\theta$:
\begin{align*}
    u(\theta)(t) = \sum_{j=1}^J b_{0,j}\mathbf{1}(c_{0,j-1} \leq t \leq c_{0,j}),\;
    u:[0,1]\to\R,\;
    \theta \equiv (b_{0,1},\ldots,b_{0,J},c_{0,1},\ldots,c_{0,J-1}).
\end{align*}
The endpoints of the support ($c_0\equiv0$, $c_J\equiv1$) and the integer $J$ are known, and coefficients $b_j$ must be positive. The true data-generating model is given by
\begin{align*}
    y = K\tilde{u}(\theta),\;
    \tilde{u}(\theta) = \epsilon(\theta) \cdot u(\theta),
\end{align*}
where $\cdot$ indicates pointwise multiplication. Here, $\epsilon(\theta)$ is a random function dependent on $\theta$ and having log-normal coefficients:
\begin{align*}
    \epsilon(\theta)(t) = \sum_{j=1}^J \epsilon_j \mathbf{1}(c_{j-1}\leq t\leq c_j),\;
    {\epsilon_j\stackrel{iid}{\sim}\mathrm{LogNormal}(-0.5,1).}
\end{align*}
{Note that each $\epsilon_j$ has distribution chosen such that the mean is 1 so that the mean of $\tilde{u}(\theta)$ is $u(\theta)$.}

We consider a linear, bounded operator $K$ {that arises in geophysical application \citep{hansen2010discrete} to define the forward model:}
\begin{align*}
    (Kv)(t) = \int \frac{1}{\{ 1 + (s-t)^2 \}^{3/2}}~v(s)ds.
\end{align*}
We observe $n=90$ independent vectors $y_1,\ldots,y_n\in\R^d$, which are pointwise evaluations of simulated curves $K\tilde{u}_i(\theta)$ on a fixed grid in [0,1]. The goal is to identify parameter $\theta$. 
The unregularized problem is ill-posed due to the smoothing nature of $K$.  
We fixed $J=4$ in the experiments, dividing up the interval into three subintervals. {The true function $u(\theta)\equiv 4$, meaning there is no unique ``optimal cutpoints'' $c_1,\ldots,c_{J-1}$}.
Assuming we do not have knowledge of this mechanism, we solve the variational problem of the form \eqref{gibbs_problem} {
\begin{equation*}
    L(\theta,y) := ||L(\theta) - y||_{\ell_1} = \sum_{i=1}^{d}|\F(\theta)_i - y_i|.
\end{equation*}
}
We place independent {Gamma(2,1)} priors on the coefficients and independent Uniform[0,1] priors on the endpoints of subregions under ordering constraints. 
{The loss is normalized by $\hat{W}_0$, an empirical estimator derived as
\begin{equation*}
    \hat{W}_0 = \frac{1}{2\hat{s}}
\end{equation*}
for a suitable scale estimator from the data. The motivation is, of course, the resemblance of the form \eqref{gibbs_formula} to a Gaussian density. 
In this problem, we plug in as $\hat{s}$ the geometric mean of sample variances across different values of $t\in[0,1]$.
}

As a benchmark, we want to compare the quality of Gibbs posterior estimate to a Bayes posterior related to $L$, in the sense that it is derived by ``believing $L$ is true'' and modifying it into a proper log-likelihood. {The $\ell^1$-loss can be related to the double exponential distribution with scale parameter $\hat{W}_0$. Since no further regularization parameter is introduced in the Bayesian framework, the compared Bayes estimate will simply fix $W\equiv 1$. The Gibbs estimate, on the other hand, will perform additional calibration of $W$ as mentioned in Section \ref{section3}.}

We used the particle filter algorithm to draw $S=10,000$ samples from both Gibbs and Bayes posteriors. For the Bayes posterior, we no longer need a separate calibration for $W$, so we discarded all samples from measures with $W < 1$. {Other than this, the sampling method remains essentially the same. A grid of $[2^{-8},2^{-7},\ldots,2^{-1},1]$ was used for searching $W$.} 
Figure \ref{fig_toy} (right) plots summaries for the resulting two distinct posterior solutions. {The plots include both the mean estimates from the two posteriors and the upper and lower bounds for 90\% credible region based on the quantiles of the posterior distributions}. We note that the Bayesian credible interval includes very irregular functions relative to the truth, due to {high variability in the coefficient estimates. 
Its high probability region is also much wider than that of the Gibbs posterior, so it overstates uncertainty and is ``inefficient'' with respect to the truth. Gibbs posterior high probability region contains functions essentially constant. This is essentially due to the fact that the LOOCV criterion chooses the smallest $W=2^{-8}$ in our grid and the prior exerts strong regularization effects. 
Given large degrees of model misspecification in this problem, choosing small $W$ and ``not learning too much'' indeed appears to provide more robust inverse estimates.
}

One can argue that in practice, researchers should immediately discard the $\ell^1$ function and intuit the irreducible uncertainty about the coefficient signs themselves. While technically true, this argument does not invalidate the main point of this example: Bayes posteriors can lead to spurious uncertainty quantification when the likelihood is a very poor approximation of the data-generating process. The benefits that accrue with the Gibbs posterior framework can be large, and the sampling algorithm remains a straightforward extension of that for Bayes posteriors.

\subsection{Stochastic Parameter Inversion using Waveguide Models} \label{section5_2}

\begin{figure}[t]
    \centering
    \includegraphics[width=2.5in]{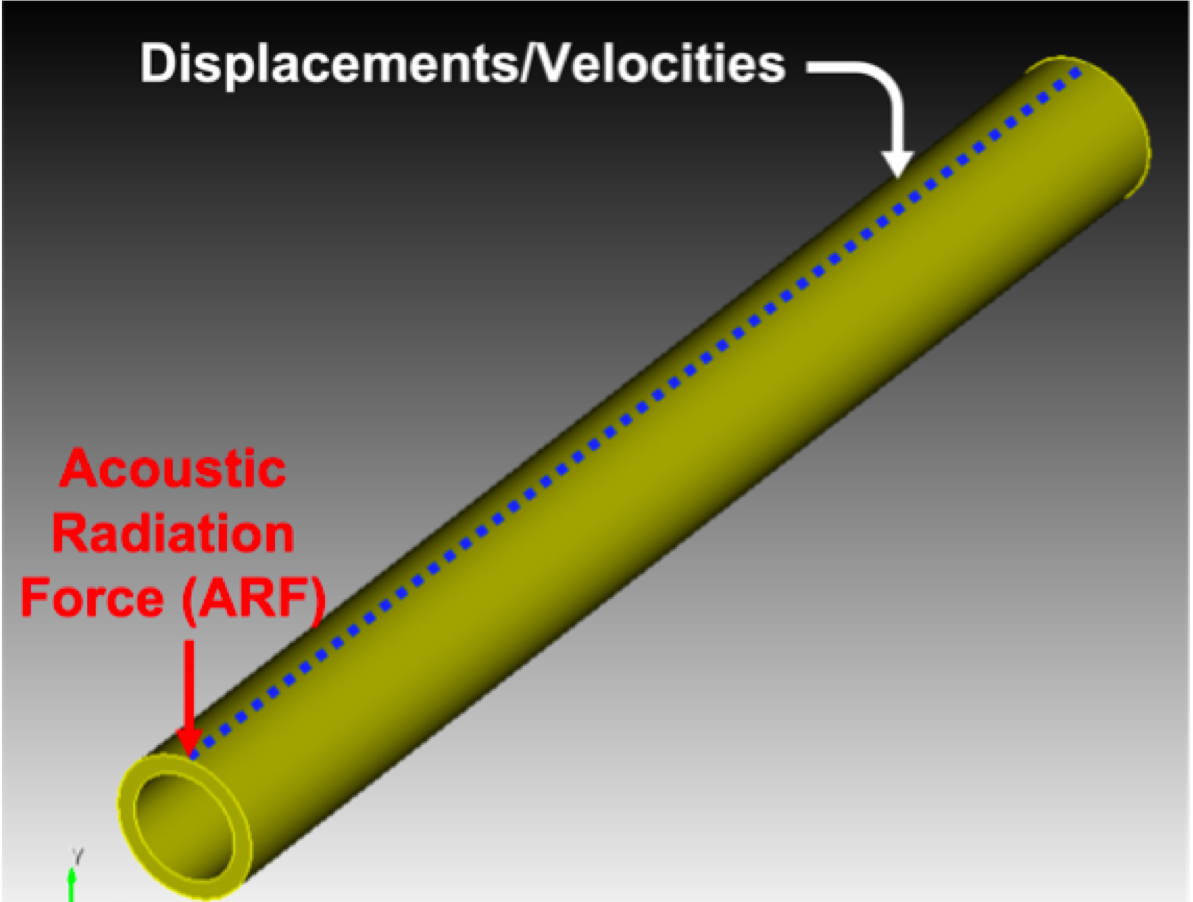}
    \caption{Waveguide with excitation and measurement locations.}
    \label{waveguide}
\end{figure}

 \begin{figure}[t]
    \centering
    \includegraphics[width=4in]{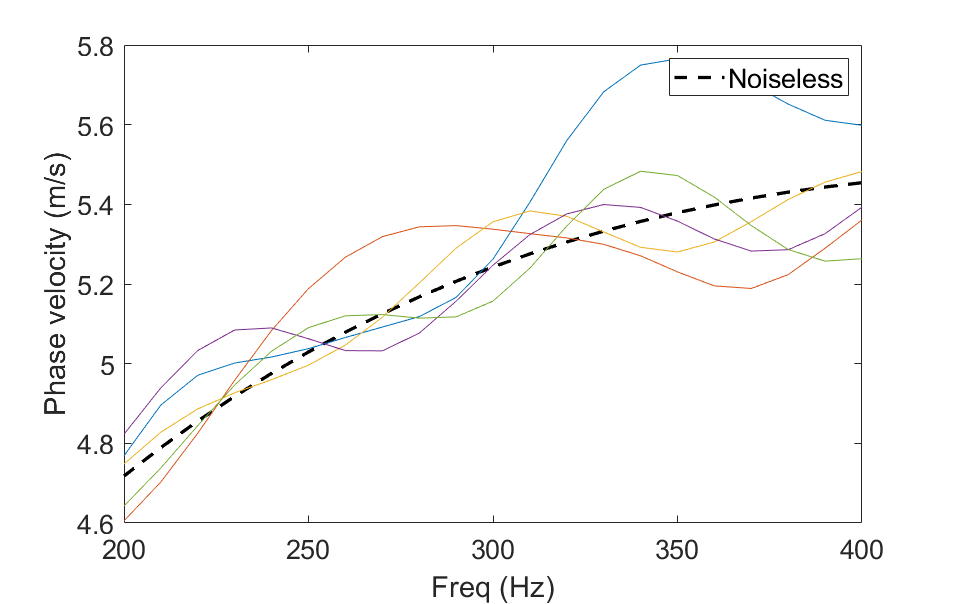}
    \caption{Simulated noisy dispersion curves for the experiment in Section \ref{section5_2} ($n=5$).}
    \label{fig_data}
\end{figure}

We turn to a simulated experiment that has more immediate relevance to practical applications. Our problem consists in estimating the material properties and geometry of a cylindrical waveguide using observations of the wave speed of propagating modes. 
This type of problem arises in numerous applications, such as non-destructive evaluation of pipes and  ultrasound elastography of arteries \citep{rm1,rm2}. Figure \ref{waveguide} shows a waveguide configuration typically found in ultrasound-based dispersion vibrometry \citep{bernal2011material,hugenberg2021toward,Capriotti_2022}.

The description of the governing equations for elastic and viscoelastic waveguides can be found in \cite{rm1} and \cite{rm2}. 
We focus on wave propagation in isotropic, linear elastic media with an unknown shear modulus. We treat the density and bulk modulus as known, a realistic assumption for nearly incompressible materials like soft tissues. 
Since we deal with cylindrical waveguides, the geometry is completely defined by their radius and thickness. In summary, our problem is to estimate with uncertainty the shear modulus, thickness, and radius of the waveguide, given noisy waveguide observations.

We now describe the data acquisition and noise modeling for the problem at hand. 
The structure is excited with a localized force, and the resulting displacements or velocities are measured on a line along the length of the waveguide, as shown in Figure \ref{waveguide}. Let us assume that these observations (i.e., velocities) can be described by the following additive noise model \eqref{add_model}:
\begin{equation} \label{add_model}
    y_i = x(\theta) + \epsilon_i,\;\epsilon_i\stackrel{iid}{\sim} P_e,\; \E\epsilon_1 = 0.
\end{equation}
Here, $\theta$ is the unknown material parameter plus geometry, and $x(\theta)$ is the particle velocity field. 
To use waveguide models, we convert the spatiotemporal observations to dispersion relations (phase velocities as functions of frequency). This step entails both a 2D Fourier transform of the data and highly nonlinear peak-finding operations. Readers interested in the details of this step are referred to \cite{bernal2011material}. 

The nonlinear transformation described maps the spatiotemporal variable $x(\theta)$ to discrete phase velocity curve $y$. 
{We can formalize this by introducing a new processing operator $\mathcal{G}$ to the additive noise model \eqref{add_model}:
\begin{equation} \label{wg_model}
    y_i = \mathcal{G}(x(\theta) + \epsilon_i) =: \F(\theta) + \delta_i(\theta),
\end{equation}
where $\F$ is the waveguide model and i.i.d. errors $\delta_i$ are potentially dependent on $\theta$. Even when space-time domain errors $\epsilon_i$'s are Gaussian,}
specifying a closed-form likelihood for \eqref{wg_model} remains a challenge, as a formal Taylor expansion shows that $\delta_i(\theta)$ depends nonlinearly on $\theta$ and random variables $\epsilon_i$s. Hence, the type of problem considered in this Section provides a prime example for {model misspecification} and the utility of Gibbs posterior framework.

To emphasize the methodological aspects of the problem, we will use simulated data, instead of \textit{in vivo} experimental data, to study the performance of Gibbs posteriors. To this end, we generated data samples by simulating the physical response through a waveguide model, using fixed ground truth parameter $\theta_0\in(\R^+)^3$, and evaluated the output of our forward waveguide model $\F$.
The first dominant mode is then perturbed by random noise. We used an independent multiplicative noise process for each of the simulated $n=5$ sample curves, following a log-Gaussian process in the frequency domain (see Figure \ref{fig_data}).

We hypothesize that prior information for the possible range of $\theta$ is available based on physical knowledge, but different notions can exist for measuring the misfit between the inverse solution and the data. 
{We consider two different losses already considered by \cite{rm2}:
\begin{itemize}
    \item Squared $\ell^2$-norm loss:
    $L_{ls}(\theta,y) := ||\F(\theta)-y||^2;$
    \item $\ell^1$ error loss:
    $L_{l1}(\theta,y) := ||\F(\theta)-y||_{\ell^1}.$
\end{itemize}
Model selection rule proposed in Section \ref{section2_4} can be used to compare the associated predictive accuracy of the two losses. Furthermore, for a meaningful comparison, we have scaled the losses based on a scale estimate from the data,
$\hat{W}_0$, based on the geometric mean of sample variances of $y_i$'s across frequencies. This turns out to be $\approx 41$.
} 

\begin{figure}[t]
    \centering
    \includegraphics[width=6in]{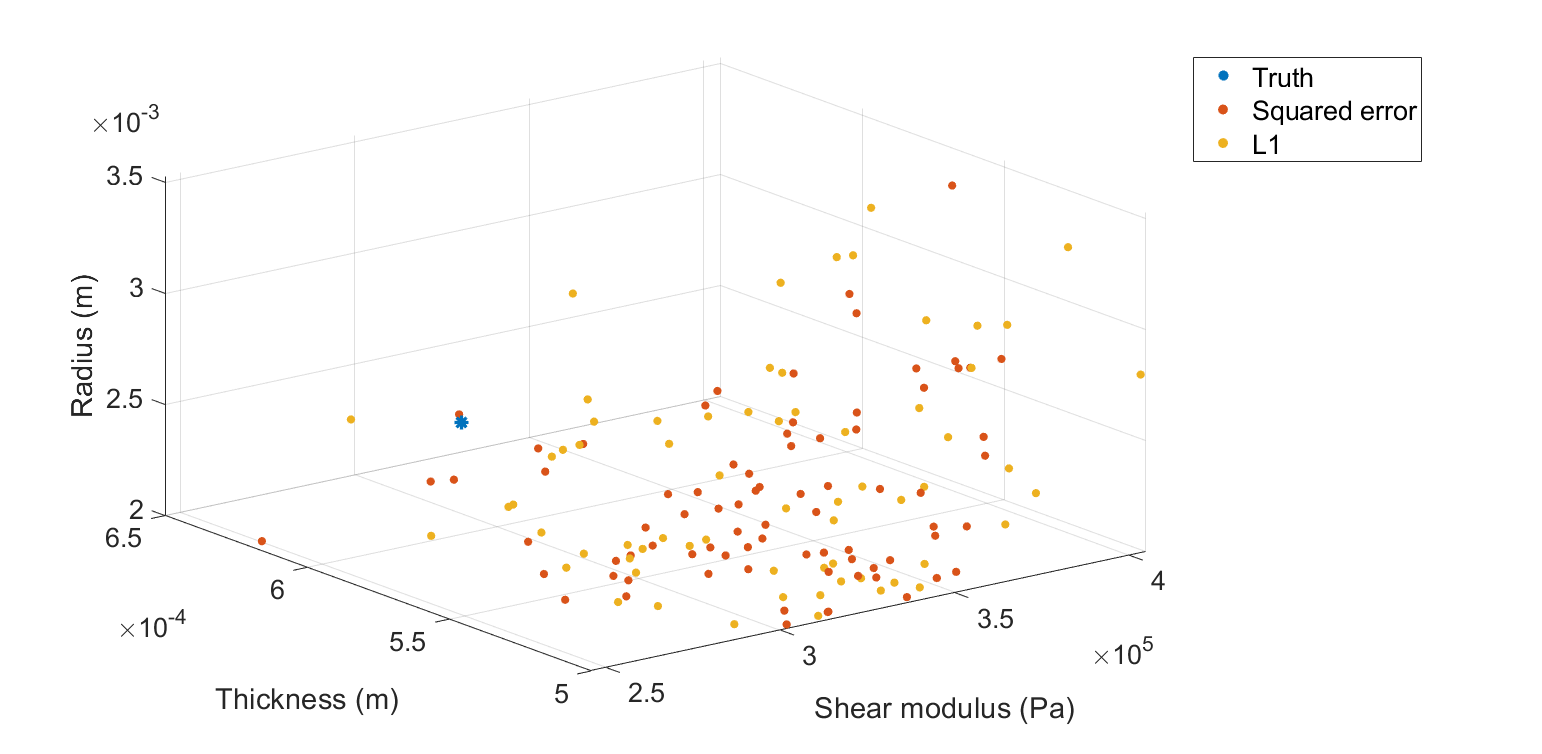}
    \caption{Joint comparison of Gibbs posterior sample draws for $\theta$ using losses $L_{ls}$ and $L_{l1}$.}
    \label{fig_posterior1}
\end{figure}

\begin{figure}[t]
    \centering
    \includegraphics[width=5.5in]{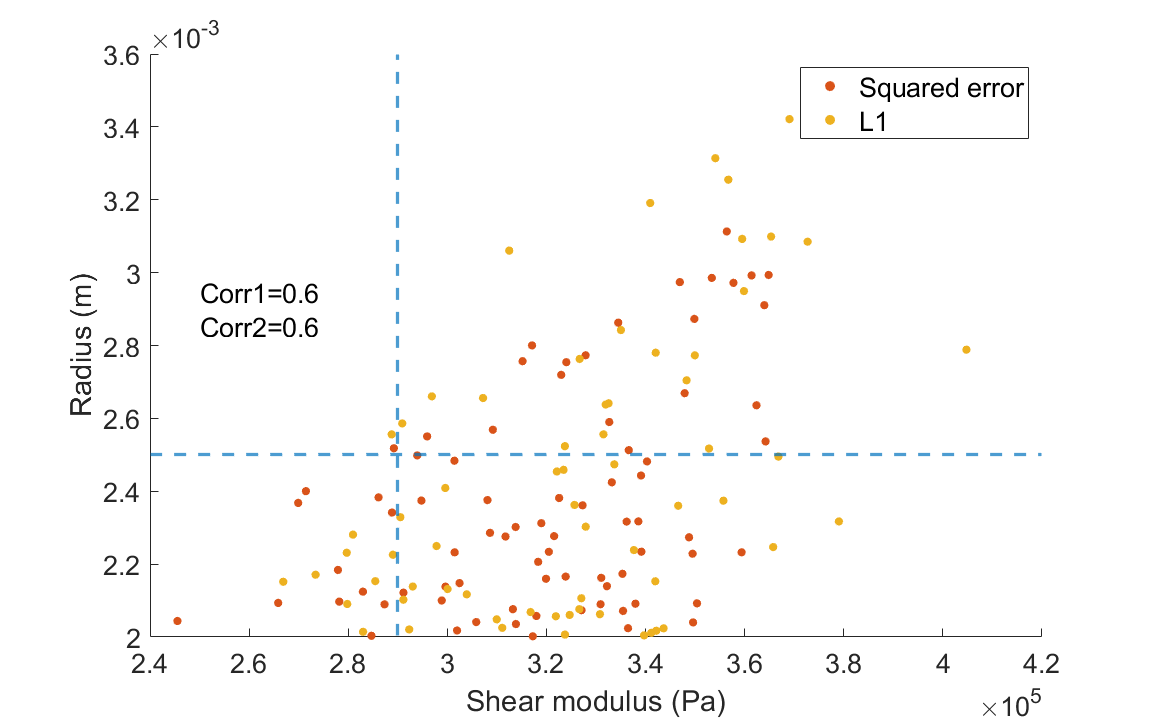}
    \caption{2D projection of Figure \ref{fig_posterior1} onto radius on shear modulus axes. Large positive correlations are induced \emph{a posteriori} (see legends).}
    \label{fig_posterior2}
\end{figure}

\begin{figure}[t]
    \centering
    \includegraphics[width=7in]{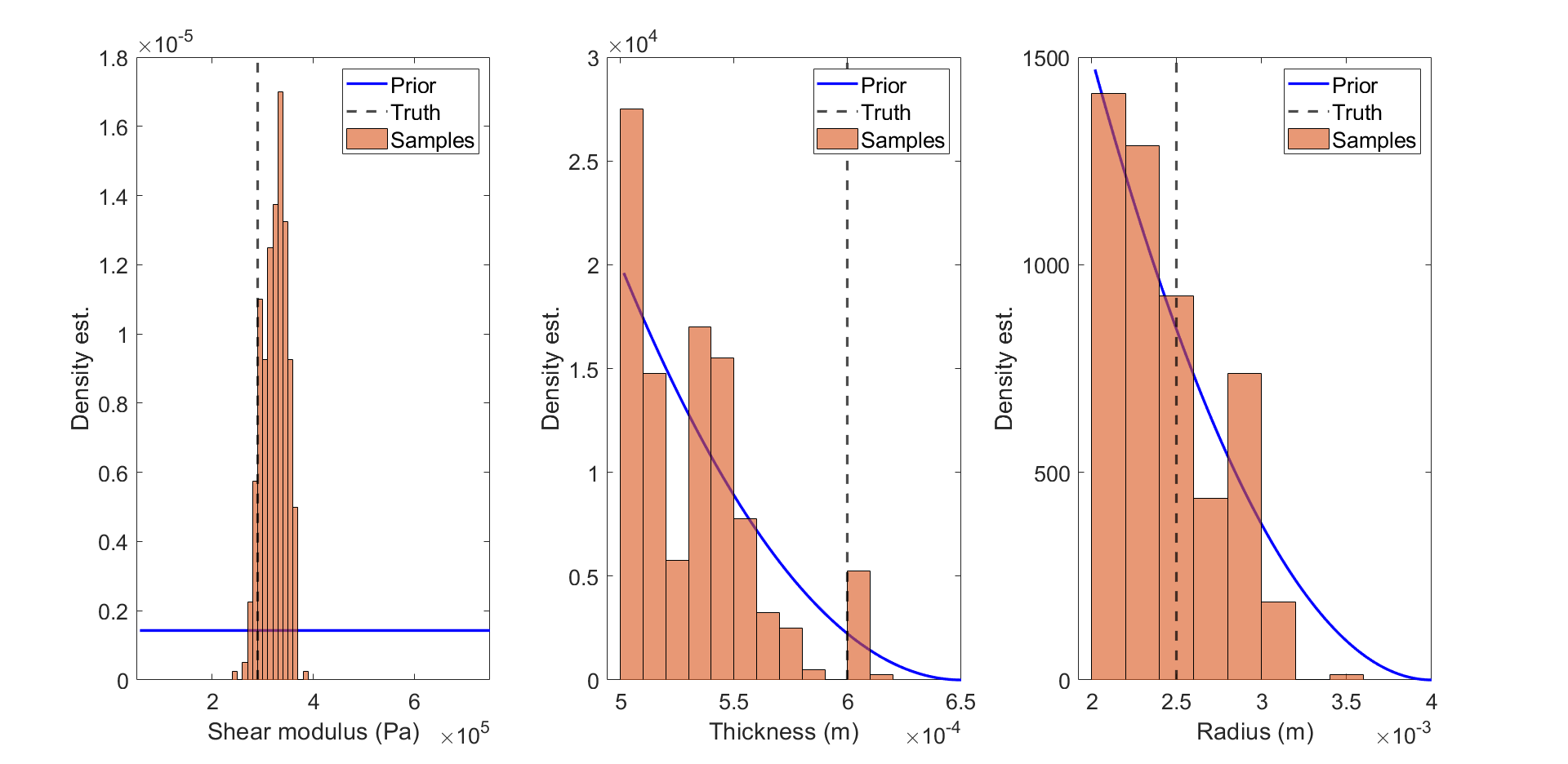}
    \caption{Marginal comparison of prior density against posterior sample draws for {the three model parameters}, using loss $L_{ls}$.}
    \label{fig_posterior3}
\end{figure}

\begin{table}[t]
    \centering
    \begin{tabular}{c|c|c|c|c|c|c}
    & 
    \multicolumn{3}{c|}{$L_{ls}$} & 
    \multicolumn{3}{c}{$L_{l1}$} \\
    \cline{2-7}
    &
    Modulus & Thickness & Radius
    & Modulus & Thickness & Radius \\
    \hline
    $|\theta_{0,i}-\E\theta_i|$ & 
    0.13 & 0.11 & 0.11 & 0.14 & 0.11 & 0.14 \\
    \hline
    $\sqrt{Var(\theta_i)}$ & 
    0.08 & 0.04 & 0.13 & 0.09 & 0.04 & 0.15 \\
    \hline
    \end{tabular}
    \caption{{
    Table of posterior mean squared errors and standard deviations. Each quantity is scaled by the magnitude of parameter $|\theta_{0,i}|$ used for simulation.
    }}
    \label{tab1}
    
    \centering
    \begin{tabular}{c|c|c}
    & $L_{ls}$ & $L_{l1}$  \\
    \hline
    $W$ & $2^{-5}$ & $2^{-6}$ \\\hline
    $P_{CV}$ & -23 & 3  \\\hline
    $\mathrm{SE}$ & -2 & 18 \\\hline
    \end{tabular}
    \caption{
    {
    Table of calibrated parameters for each model among a grid $[2^{-10},2^{-9},\ldots,2^{-1},1]$. Lower $P_{CV}$ \eqref{pred_criterion} implies relative predictive optimality. $\mathrm{SE}$ is the approximate standard error of $P_{CV}$.
    }}
    \label{tab2}
\end{table}

An independent ${\rm Beta}(1,3)$ prior was placed on each component of $\Theta$. The prior was translated and re-scaled to have support on the interval dictated by physical knowledge for each parameter. The intervals were between 5 and 95~kPa for shear modulus, between 5~mm and 6.5~mm for arterial wall thickness, and between 2~mm and 4~mm for artery radius, respectively. We used the SMC algorithm of Section \ref{section3_3} to draw $S=400$ samples from all four Gibbs posteriors. In successive iterations, particles were mutated by running a Metropolis-Hastings algorithm for $K=5$ steps. 

Figure \ref{fig_posterior1} plots two distinct sets of posterior sample draws obtained after terminating the algorithm. Apparently, both $L_{ls}$ and $L_{l1}$ lead to posteriors characterized by similar high probability regions. {Table \ref{tab1} vindicates this based on mean squared errors of the posterior mean estimators (accuracy) and posterior standard deviations (nominal uncertainty) for each parameter.} For both posteriors, $W$ chosen through LOOCV was less than 1. Figure \ref{fig_posterior2} shows that both distributions also exhibit a positive linear correlation between shear modulus and radius parameters, 
 a structure learned as opposed to an independent prior specified on the parameters.

Figure \ref{fig_posterior3} demonstrates the difficulty of inferring simultaneously material and geometric parameters in waveguides. Here, histograms of posterior sample draws are compared against the prior densities. For the thickness and radius parameters, posterior density estimates closely follow the prior density
function. This implies little learning has taken place for these parameters marginally. {On the other hand, the marginal density estimate for the shear modulus parameter is highly concentrated on a neighborhood of the true value used in simulation, in contrast with a uniform prior on the support.} 
 While the Figure shows only the posterior for $L_{ls}$, a similar lack of learning took place for $L_{l1}$, indicating a lack of information in the data and the need for stronger prior knowledge to distinguish the effect of using different loss functions. Practically, one can resolve the problem by collecting richer data with more propagating modes and different frequency ranges.

 Table \ref{tab2} lists the statistics $P_{CV}$ for the {two losses being compared. \cite{rm1} have remarked that empirically, there seems to be little difference in choosing either loss. From the point of view of predictive criterion we suggest, on the other hand, using $L_{ls}$ yields a much smaller statistic $P_{CV}$ on average.}

\section{Discussion} \label{section6}

In this work, we have justified using Gibbs posteriors as probabilistic extensions of regularized estimators in classical inverse problems. These probabilistic solutions possess important philosophical and theoretical advantages over traditional Bayes posteriors in model misspecification scenarios. We have introduced numerical methods to implement model calibration and selection. We also demonstrated practical applications of our methods through simulated experiments.

As observed by many authors \citep{owhadi2015brittleness,owhadi2017qualitative}, the adoption of a strictly Bayesian approach to inverse problems can lead to various fundamental questions. The complexity of the forward model and the parameter space pose challenges like extreme sensitivity to prior selection, model perturbations, and numerical approximation errors. Another important challenge is to rigorously formulate the adequacy of a given uncertainty quantification. \cite{kleijn2012bernstein}, for example, studied the frequentist coverage and Bernstein-von Mises phenomena for misspecified Bayes posteriors, with mixed results. The variational formulation of our paper, as it stands now, does not address these issues. Whether the Gibbs posterior framework can allow us to overcome these challenges is an interesting question. A promising approach has been laid out by \cite{sm19}, who proposed the GPC algorithm to choose $W$ that explicitly meets frequentist coverage demands.

Another important future task is to extend our framework to a fully nonparametric setup, where the parameter space is now infinite-dimensional. Despite the increasing number of works in this area, there remains a lot to be answered about the properties of misspecified posteriors on infinite dimensional spaces. Questions regarding consistency and frequentist coverage of the posteriors also become even more delicate \citep{castillo2013nonparametric,castillo2014bernstein}. We believe adopting a loss-based framework with a clever choice of regularization parameter $W$ is a promising avenue for resolving many difficulties in nonparametric inverse problems.

\section*{Acknowledgements}
Sayan Mukherjee would like to acknowledge partial funding from HFSP RGP005, NSF
DMS 17-13012, NSF BCS 1552848, NSF DBI 1661386, NSF IIS 15-46331, NSF DMS 16-13261, as well as high-performance computing partially supported by grant 2016-IDG-1013
from the North Carolina Biotechnology Center. Wilkins Aquino would like to acknowledge partial funding from NIH-R01HL145268-04. All authors thank Dr. Tuhin Roy and Prof. Murthy Gudatti, Department of Civil and Environmental Engineering, North Carolina State University for allowing the use of their waveguide simulation code.

\newcommand{\newblock}{}
\bibliography{inverse_paper_rev1}

\appendix

\section{Proofs} 
\label{section_A}

\subsection*{Proof of Theorems \ref{thm_stability}, \ref{thm_approx}}

We first prove Theorem \ref{thm_stability}. Our proof uses essentially the same estimates as \cite{stuart2010} in his Theorem 4.2, but in different ways. 

\begin{proof}
    First, we note the following holds for the two normalizing constants of $\hrho_{n,1}^W$ and $\hrho_{n,2}^W$ due to assumptions \ref{a_loss}.\ref{lower} and \ref{a_loss}.\ref{lipschitz_data}:
    \begin{align*}
        |Z_{n,1}^W - Z_{n,2}^W| &\leq
        \int |e^{-W\sum_{i=1}^nL(\theta,y_{1,i})} - e^{-W\sum_{i=1}^{n}L(\theta,y_{2,i})}|~\rho_0(\theta)d\theta \\
        &= \int e^{-W\sum_{i=1}^{n}L(\theta,y_{1,i})}
        W\sum_{i=1}^{n}|L(\theta,y_{1,i}) - L(\theta,y_{2,i})|~\rho_0(\theta)d\theta \\
        &\leq W\int C_2(r,\theta)\sum_{i=1}^n ||y_{1,i}-y_{2,i}||~\rho_0(\theta)d\theta \\
        &=: W\tilde{C}(r)\sum_{i=1}^{n}||y_{1,i} - y_{2,i}||.
    \end{align*}
    Next, by the integrability condition \ref{a_loss}.\ref{upper} and Jensen's inequality, there exists $K(r)$ such that
    \begin{align*}
        Z_{n,1}^W &\geq\exp\left\{-W\sum_{i=1}^{n}\int L(\theta,y_{1,i})~\rho_0(\theta)d\theta\right\}\geq e^{-nWK(r)}.
    \end{align*}
    Combining these estimates and the definition of KL-divergence, we obtain the following:
    \begin{align*} 
        \kl(\hrho_{n,1}^W||\hrho_{n,2}^W) &=
        \int\left[-W\sum_{i=1}^{n}L(\theta,y_{1,i}) + W\sum_{i=1}^{n}L(\theta,y_{2,i}) - \log Z_{n,1}^W + \log Z_{n,2}^W\right]~\hrho_{n,1}^W(\theta)d\theta \\
        &\leq (Z_{n,W})^{-1}\int\left[WC_2(r,\theta)\sum_{i=1}^{n}|L(\theta,y_{1,i} - L(\theta,y_{2,i}| + |\log Z_{n,1}^W - \log Z_{n,2}^W|\right]~\rho_0(\theta)d\theta\\
        &\leq (Z_{n,1}^W)^{-1}\left[W\tilde{C}(r)\sum_{i=1}^{n}||y_{1,i} - y_{2,i}|| + \log\left(\frac{Z_{n,1}^W\vee Z_{n,2}^W}{Z_{n,1}^W\wedge Z_{n,2}^W}\right)\right] \\
        &\leq(Z_{n,1}^W\wedge Z_{n,2}^W)^{-1}\left[W\tilde{C}(r)\sum_{i=1}^{n}||y_{1,i} - y_{2,i}|| + \log\left(\frac{Z_{n,1}^W\vee Z_{n,2}^W}{Z_{n,1}^W\wedge Z_{n,2}^W}\right)\right],
    \end{align*}
    where 
    \begin{align*}
        \log\left(\frac{Z_{n,1}^W\vee Z_{n,2}^W}{Z_{n,1}^W\wedge Z_{n,2}^W}\right) &\leq
        \log\left(1 + \frac{W\sum_{i=1}^{n}\tilde{C}(r)||y_{1,i} - y_{2,i}||}{Z_{n,1}^W\wedge Z_{n,2}^W}\right) \\
        &\leq (Z_{n,1}^W\wedge Z_{n,2}^W)^{-1}W\tilde{C}(r)\sum_{i=1}^{n}||y_{1,i} - y_{2,i}||.
    \end{align*}
    Exchanging the positions of two measures yields the same bound on $\kl(\hrho_{n,2}^W||\hrho_{n,1}^W)$. We use the lower bound on $Z_{n,1}^W$ and $Z_{n,2}^W$, and the fact that $x(1+x)\leq (1+x)^2$ for $x\geq 0$, to deduce there exists a choice $K(r)$ such that
    \begin{equation*}
    \kl(\hrho_{n,1}^W||\hrho_{n,2}^W)\vee
    \kl(\hrho_{n,2}^W||\hrho_{n,1}^W) 
    \leq
    W\tilde{C}(r)(1 + e^{nWK(r)})^2\sum_{i=1}^{n}||y_{1,i}-y_{2,i}||^2.
    \end{equation*}
    Setting $a=\tilde{C}(r)$ and $b=K(r)$, the claimed upper bound on KL divergence follows.

    The following inequality due to \cite{bretagnolle1979estimation} yields the claimed upper bound on the Hellinger (and total variation) distance:
    \begin{equation} \label{bh_inequality}
    d_{H}^2(\hrho_{n,1}^W,\hrho_{n,2}^W) \leq d_{TV}(\hrho_{n,1}^W,\hrho_{n,2}^W) \leq \sqrt{1-\exp(-\{\kl(\hrho_{n,1}^W||\hrho_{n,2}^W)\wedge\kl(\hrho_{n,2}^W||\hrho_{n,1}^W)\})}.
\end{equation}
Hence, Theorem \ref{thm_stability} is proven.
\end{proof}

The proof of Theorem \ref{thm_approx} follows a very similar argument, so we do not repeat all the details. In brief, under the stated assumptions of \ref{thm_approx}, there exists some $\tilde{C}(r)$ such that:
\begin{align*}
    |Z_n^W - Z_{n,h}^W| &\leq \int|e^{-W\sum_{i=1}^{n} L(\theta,y_i)} - e^{-W\sum_{i=1}^{n} L^h(\theta,y_i)}|~\rho_0(\theta)d\theta\\
    &\leq W\tilde{C}(r)\sum_{i=1}^{n}||y_{1,i} - y_{2,i}||,
\end{align*}
and there exists $K(r)$ such that:
\begin{align*}
    \kl(\hrho_n^W||\hrho_{n,h}^W) &\leq (Z_n^W)^{-1}
    \int\left[W\tilde{C}(r)\psi(h) + |\log Z_n^W - \log Z_n^W|\right]~\rho_0(\theta)d\theta\\
    &\leq (Z_n^W\wedge Z_{n,h}^W)^{-1}\left[W\tilde{C}(r)\sum_{i=1}^{n}\psi(h) + \log\left(\frac{Z_n^W\wedge Z_{n,h}^W}{Z_n^W\vee Z_{n,h}^W}\right)\right]\\
    &\leq e^{nWK(r)}(1+e^{nWK(r)})W\tilde{C}(r)\psi(h)\\
    &\leq W\tilde{C}(r)(1+e^{nWK(r)})^2\psi(h).
\end{align*}
Again, due to \eqref{bh_inequality}, Theorem \ref{thm_approx} follows.

\subsection*{Proof of Theorem \ref{thm_consistency}}

\begin{proof}
    Many of our arguments here are routine steps in posterior consistency literature, first put forth by \cite{schwartz1965bayes}. First, we show there exist a sequence of bounded functions $\phi_n:\Y^n\to\{0,1\}$ satisfying, for every $\epsilon>0$:
    \begin{align*}
        \int\phi_n(y_1,\ldots,y_n)~\prod_{i=1}^{n}\P(y_i)dy_i &\to 0,\\
        \sup_{||\theta-\theta^*||>\epsilon}\int\{1-\phi_n(y_1,\ldots,y_n)\}
        \prod_{i=1}^{n}e^{-W\{L(\theta,y_i)-L(\theta^*,y_i)\}}&\P(y_i)dy_i\to 0.
    \end{align*}
    This sequence provides an analogue to Schwartz's test functions \citep{schwartz1965bayes} and was first considered by \cite{kleijn2012bernstein}. In particular, since we assume $\Theta$ is compact, by Theorem 3.2 \citep{kleijn2012bernstein} it is enough to show for existence that the maps
    \begin{equation*}
        \theta\mapsto\exp[-W\{L(\theta,y) - L(\theta',y)\}]
    \end{equation*}
    and
    \begin{equation*}
        \theta\mapsto\exp[-W\{L(\theta,y) - L(\theta^*,y)\}]
    \end{equation*}
    are continuous at $\theta = \theta'$ for every $\theta'\in\Theta$. The first continuity is an easy consequence of Assumption \ref{a_loss}.\ref{lipschitz_par}. The second continuity follows also from Assumption \ref{a_loss}.\ref{lipschitz_par} and the additional assumption that $C_1(y)e^{\bar{W}L(\theta^*,y)}\in L^1(\Y;\P)$, so as extension for every $W<\bar{W}$. Finally, by Lemma 3.3 \citep{kleijn2012bernstein} we can choose $\phi_n$ such that the rate of convergence is exponentially fast.

    Let $U$ be an arbitrary open set containing $\Theta^*$. We now show that $\hrho_n^W(\Theta\setminus U)\to 0$ with probability 1 by rewriting, for any choice of $\theta^*\in\Theta^*$:
    \begin{equation*}
        \hrho_n^W(\Theta\setminus U) \leq \underbrace{\phi_n}_{(*)} +
        \underbrace{\frac{\int_{\Theta\setminus U}(1-\phi_n)\exp[-W\sum_{i=1}^{n}\{L(\theta,y_i) - L(\theta^*,y_i)\}]\rho_0(\theta)d\theta}{\int\exp[-W\sum_{i=1}^{n}\{L(\theta,y_i) - L(\theta^*,y_i)\}]~\rho_0(\theta)d\theta}}_{(**)}.
    \end{equation*}
    There exists $C,D>0$ such that $\mathbb{E}[(*)]$ is less than $e^{-nC}$ and the expectation of the numerator of (**) is less than $e^{-nD}$. It remains to control the denominator of (**). First, note that by assumption \ref{a_prior}, for arbitrary $\epsilon>0$,
    \begin{equation*}
        \rho_0\{\theta:||\theta-\theta^*||<\epsilon\} \geq
        \inf_{\theta:||\theta-\theta^*||<\epsilon}\rho_0(\theta) \geq
        \rho_0(\theta^*)e^{-C_3\epsilon} > 0.
    \end{equation*}
    Hence, the prior probability $\rho_0\{\theta:||\theta-\theta^*||<\epsilon\}$ is bounded away from zero.
    Using Jensen's inequality and Assumption \ref{a_loss}.\ref{lipschitz_par}, for arbitrary $\epsilon$, we obtain a finite lower bound
    \begin{align*}
        \log\int\exp\left[-nW\{R_n(\theta) - R_n(\theta^*)\}\right]~\rho_0(\theta)d\theta &\geq \log\int_{||\theta-\theta^*||<\epsilon}
        \exp\left[-nW\{R_n(\theta) - R_n(\theta^*)\}\right]~\rho_0(\theta)d\theta \\
        &\geq -nW\int_{||\theta-\theta^*||<\epsilon} \{R_n(\theta) - R_n(\theta^*)\}~\rho_0(\theta)d\theta \\
        &\geq -\epsilon W\sum_{i=1}^{n}C(y_i).
    \end{align*}
    By strong law of large numbers, $\frac{1}{n}\sum_{i=1}^{n}C(y_i)\to c$ for some $c$ with probability 1. Therefore, for every $c'>c$, eventually $n$ becomes large enough such that the denominator of (**) is at least $e^{-nc'\epsilon}$.
    Since choice of $\epsilon$ is arbitrary, let it be small enough such that $c'\epsilon < D$. Then,
    \begin{equation*}
        \E\left[e^{nc'\epsilon}
        \int_{\Theta\setminus U}(1-\phi_n)\exp[-W\sum_{i=1}^{n}\{L(\theta,y_i) - L(\theta^*,y_i)\}]\rho_0(\theta)d\theta\right]
        \leq e^{-n(D-c'\epsilon)}\stackrel{n}{\to} 0.
    \end{equation*}
    By using Markov's inequality and Borel-Cantelli theorem, we deduce that the term inside the expectation converges to 0 with probability 1. Since this term is greater than (**) for large $n$, we conclude that $\hrho_n^W(\Theta\setminus U)\to 0$ with probability 1.
\end{proof}

\subsection*{Proof of Theorem \ref{thm_pred_const}}

\begin{proof}
We consider the convergence of quantity
\begin{equation*}
    -\int\log\frac{\hat{p}_n^W(\ynew)}{\tilde{p}_{\theta^*}}~\P(\ynew)d\ynew.
\end{equation*}
Using the definition of the densities involved, this equals
\begin{align*}
    &\overbrace{\int\left(-\log\int e^{-L(\theta,\ynew)}~\hrho_n^W(\theta)d\theta - L(\theta^*,\ynew)\right)~\P(\ynew)d\ynew}^{(*)}
    \\
    &\quad\quad+\underbrace{\log\iint e^{-L(\theta,y')}~\hrho_n^W(\theta)d\theta~\lambda(y')dy' - \log\int e^{-L(\theta^*,y')}~\lambda(y')dy'}_{(**)}.
\end{align*}
Exchanging the order of integration by Fubini's theorem and rewriting the integrals in terms of measures, (**) equals
\begin{equation*}
   \log\iint e^{-L(\theta,y')}~\lambda(y')dy'~\hrho_n^W(\theta)d\theta - \log\iint e^{-L(\theta,y')}~\lambda(y')dy'~\delta_{\theta^*}(d\theta).
\end{equation*}
By assumptions \ref{lower} and \ref{lipschitz_par}, $\int e^{-L(\theta,y')}~\lambda(y')dy'$ is bounded and continuous in $\theta$. By definition of weak convergence for probability measures, we conclude that (**) converges to 0 with probability 1. 

For (*), a similar argument shows that for every fixed $\ynew$, the difference
\begin{equation*}
    -\log\int e^{-L(\theta,\ynew)}~\hrho_n^W(\theta)d\theta - L(\theta^*,\ynew)
\end{equation*}
converges to 0 with probability 1. Fatou's lemma can be used to yield a lower bound that converges to 0, and Jensen's inequality to yield an upper bound:
\begin{align*}
    \int &\underbrace{\lim_{n\to\infty}\left\{-\log\int e^{-L(\theta,\ynew)}~\hrho_n^W(\theta)d\theta - L(\theta^*,\ynew)\right\}}_{\to 0} \P(\ynew) \\ &\leq (*)\\
    &\leq \iint \{L(\theta,\ynew) - L(\theta^*,\ynew)\}~\hrho_n^W(\theta)d\theta~\P(\ynew)d\ynew
\end{align*}
Under assumption of Theorem \ref{thm_consistency}, $\Theta$ is compact. This implies $||\theta-\theta^*||\leq B$ for a uniform bound $B$. By assumption \ref{a_loss}.\ref{lipschitz_par}, $L(\theta,\ynew) - L(\theta,\ynew) \leq C_1(\ynew)B$. Since $C_1\in L^1(\Y;\P)$ by assumption, by dominated convergence theorem, we conclude the upper bound for (*) also converges to 0 with probability 1.
\end{proof}

\clearpage

\section{Algorithms}
\label{section_B}

\begin{algorithm}
\caption{Particle filter sampler for calibrating $W$}
\label{alg_filter_cal}

\KwData{$\boldsymbol{y} = (y_1,\ldots,y_n)$, $S,K,E_{min},(W_1,\ldots,W_{T-1})$}
\KwResult{$W^* \in (0,1],\; 
\theta^{(1)},\ldots,\theta^{(S)}
\stackrel{approx}{\sim} \rho_{mix}^{W^*},\;
P_{CV}$}

$t\gets 0$\;
$W_T\gets 0$\;
$W\gets 0$\;
$R_{CV}\gets (0,\ldots,0)$
\While{$t<T$}{
  $t\gets t + 1$\;
  \eIf{$t=1$}{
    \For{$s=1,\ldots,S$ in parallel}{
      $\theta^{(s)} \sim \rho_0$\;
      $W\gets W_t$\;
      $w_t (\theta^{(s)})\gets
      \sum_{i=1}^{n} \exp\{-W\sum_{j:j\neq i} L(\theta,y_j)\}$\;
    }
  }{
  $w_t(\theta^{(s)})\gets\frac{\sum_{i=1}^{n} \exp\{-W_t\sum_{j:j\neq i} L(\theta,y_j)\}}
  {\sum_{i=1}^{n} \exp\{-W\sum_{j:j\neq i} L(\theta,y_j)\}}$\;
  $W\gets W_t$\;
  $ESS \gets \frac{\left(\sum_{s=1}^{S}w_t(\theta^{(s)})\right)^2}
  {\sum_{s=1}^{S}w_t(\theta^{(s)})^2}$\;
  \If{$ESS < E_{min}$}{
    $A^{1:S}\gets\text{Resample}(w_t(\theta^{(1)}), \ldots,
    w_t(\theta^{(S)}))$
    \Comment*[r]{Generic resampling algorithm}
    $\theta^{(s)} \gets \theta_{A^s}$\;
    \For{$s=1,\ldots,S$ in parallel}{
      \For{$k=1,\ldots,K$}{
        $\theta_{prev}^{(s)} \gets 
        \theta^{(s)}$\;
        $\theta^{(s)} \gets 
        \text{Transition}(\theta_{prev}^{(s)},d\theta)$ \Comment*[r]{Generic Markov transition kernel preserving invariance}
      }
      }
      }
      \For{$i=1,\ldots,n$}{
        $r_i(\theta^{(s)})\gets
        \frac{\exp\{WL(\theta^{(s)},y_i)\}}{\sum_{j=1}^{n}
        \exp\{WL(\theta^{(s)},y_j)\}}$
      }
      Compute $R_{CV}[t]$
      \Comment*{See \eqref{approx_obj_cv}}
    }
    $t^*\gets \argmin_t R_{CV}$\;
    $W^*\gets W_{t^*}$\;
    Compute $P_{CV}$ \Comment*{See \eqref{approx_pred_crt}}
  }
\end{algorithm}

\begin{algorithm}
\caption{Particle filter sampler for sampling from full posterior}
\label{alg_filter_full}

\KwData{$\y = (y_1,\ldots,y_n),\; 
W^*,\; \theta^{(1)},\ldots,\theta^{(S)}
\stackrel{approx}{\sim} \rho_{mix}^{W^*}$}
\KwResult{$\theta^{(1)},\ldots,\theta^{(S)}\stackrel{approx}{\sim} \hrho_n^{W^*}$}

$w_t(\theta^{(s)}) \gets \left[\sum_{i=1}^{n}\exp\{W^*L(\theta^{(s)},y_i)\}\right]^{-1}$\;
$A^{1:S}\gets\text{Resample}(w_t(\theta^{(1)}), \ldots, w_t(\theta^{(S)}))$\;
$\theta^{(s)} \gets \theta_{A^s}$\;
\For{$s=1,\ldots,S$ in parallel}{
    \For{$k=1,\ldots,K$}{
        $\theta_{prev}^{(s)} \gets \theta^{(s)}$\;
        $\theta^{(s)} \gets \text{Transition}(\theta_{prev}^{(s)},d\theta)$ \;
    }
}

\end{algorithm}

\end{document}